\documentclass[sigplan,nonacm]{acmart}

\usepackage{mathtools}
\usepackage{listings}
\usepackage{multirow}
\usepackage{wrapfig}
\usepackage{url}
\usepackage{xcolor}
\usepackage{enumitem}
\usepackage{capt-of}
\usepackage{booktabs}
\usepackage{subcaption}
\usepackage{algorithm}
\usepackage{algpseudocode}
\usepackage{amsmath}

\usepackage{tikz}

\newcommand{\system}{\textsc{\mbox{Nalar}}\xspace}

\newcommand{\saurabh}[1]{\textcolor{blue}{SA:#1}}
\newcommand{\aditya}[1]{\textcolor{red}{AA:#1}}

\usepackage{xspace}
\usepackage{soul}

\newcommand{\ie}{{\it i.e.}, }

\lstdefinelanguage{JSON}{
    basicstyle=\ttfamily,
    showstringspaces=false,
    breaklines=true,
    string=[s]{"}{"},
    stringstyle=\color{blue},
    numbers=left,
    numberstyle=\tiny,
    backgroundcolor=\color{gray!10}
}

\lstdefinelanguage{JavaScript}{
  keywords={typeof, new, true, false, catch, function, return, null, switch, var, if, in, while, do, else, case, break, let, const},
  keywordstyle=\color{blue}\bfseries,
  ndkeywords={class, export, boolean, throw, implements, import, this},
  ndkeywordstyle=\color{teal}\bfseries,
  identifierstyle=\color{black},
  sensitive=false,
  comment=[l]{//},
  morecomment=[s]{/*}{*/},
  commentstyle=\color{gray}\ttfamily,
  stringstyle=\color{red}\ttfamily,
  morestring=[b]',
  morestring=[b]"
}

\lstdefinestyle{customc2}{
  language=JavaScript,
  basicstyle=\ttfamily\small,
  keywordstyle=\color{blue},
  stringstyle=\color{red},
  commentstyle=\color{gray},
  numbers=left,
  numberstyle=\tiny,
  breaklines=false,
  showstringspaces=false
}

\usepackage{graphicx}
\graphicspath{ {./images/} }
\definecolor{brewerpurple}{HTML}{AF4EA3}
\definecolor{brewerblue}{HTML}{377EB8}
\definecolor{NavyBlue}{HTML}{006EB8}
\definecolor{BrickRed}{HTML}{B6321C}
\definecolor{ForestGreen}{HTML}{009B55}
\lstdefinestyle{customc2}{
    emph={update\_metrics, prune\_completed\_jobs, exec\_jobs, accept, schedule, place, AcceptAll, Fifo, Consolidated, pop\_wait\_queue, update\_cluster},
    emphstyle=\bfseries\color{NavyBlue},
    commentstyle=\color{ForestGreen}\itshape\ttfamily,
       morekeywords={yield},
    stringstyle=\color{red}\ttfamily,
    emph={[2]admission\_policy, scheduling\_policy, placement\_policy},emphstyle={[2]\bfseries\color{BrickRed}},
    language=Python,
    keywordstyle=\bfseries\color{green!40!black}
}
\usepackage{circledsteps}
\begin{document}
\title{\system: Workflow-Aware Management of Agentic Applications}
% \date{}
\author{Saurabh Agarwal}
\authornote{Both authors contributed equally to this research.}
\affiliation{%
  \institution{UT Austin}
  \country{USA}
}
\author{Marco Laju}
\authornotemark[1]
\affiliation{%
  \institution{UT Austin}
  \country{USA}
}
\author{Donghyun Son}
\affiliation{%
  \institution{UT Austin}
  \country{USA}
}
\author{Nitin Kedia}
\affiliation{%
  \institution{UT Austin}
  \country{USA}
}
\author{Myungjin Lee}
\affiliation{%
  \institution{Cisco Research}
  \country{USA}
}
\author{Jayanth Srinivasa}
\affiliation{%
  \institution{Cisco Research}
  \country{USA}
}
\author{Aditya Akella}
\affiliation{%
  \institution{UT Austin}
  \country{USA}
}
\begin{abstract}
% \saurabh{asbstract fix}
LLM-driven agentic applications automate complex, multi-step tasks, but serving them efficiently remains difficult due to heterogeneous components, dynamic model-driven control flow, long-lived state, and highly variable latencies. \system is a serving framework for agent workflows that separates workflow specification from execution while providing the runtime visibility and control needed for robust performance. \system preserves ordinary Python interface and control flow through lightweight auto-generated stubs that turn agent and tool invocations into futures carrying dependency and execution-context metadata.  A two-level control architecture combines global policy computation with local event-driven enforcement to support adaptive routing, scheduling, and resource management across evolving workflows. 
A workflow-aware KV-cache layer enables the runtime to manage cache placement and lifetime.
Together, these mechanisms enable scalable, efficient, policy-driven serving of heterogeneous agentic applications without burdening developers with orchestration logic. Across three agentic workloads, \system reduces tail latency by 34--74\%, achieves up to $3.38\times$ speedups.
% New scheduling policies can be implemented in only a few lines of code.
% \aditya{check these numbers}
\end{abstract}
\maketitle
\section{Introduction}

Agentic applications~\cite{yao2022react, ge2023openagi, hong2023metagpt,
li2024autoflow, li2025webthinker} are increasingly being used to automate
complex, multi-step tasks. They compose LLM-driven agents with tools and
other services to decompose high-level goals, invoke specialized
functionality, evaluate intermediate results, and decide what to do next.
This model is being applied in many domains such as software engineering, financial
analysis, operations, and customer support. Even as individual LLM
inference becomes more efficient, serving an agentic application remains difficult. A single user request can generate many agent and tool invocations whose 
execution and resource demands change over time. Performance and resource
efficiency therefore depend on {\em coordinating the end-to-end workflow}.

\begin{figure}[t]
    \centering
    \includegraphics[width=\linewidth]{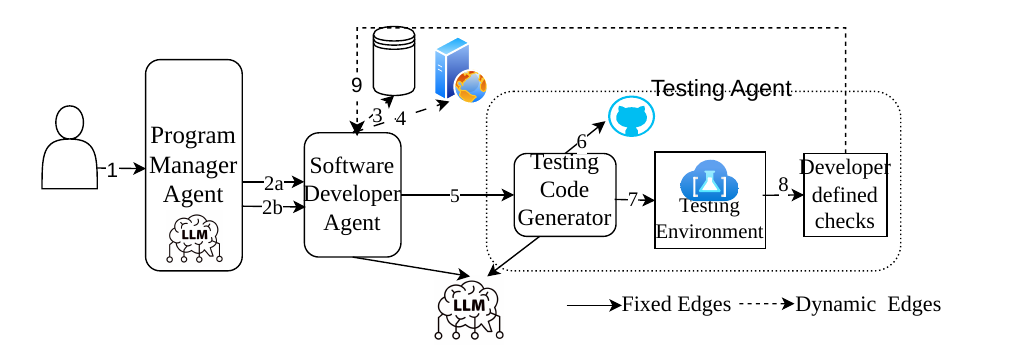}
    \vspace{-10pt}
    \caption{\textbf{An example agentic application:} OpenSWE ~\cite{open-swe} workflow for software development.}
    \label{fig:agent_workflow}
    \vspace{-15pt}
\end{figure}

Consider the simplified Open-SWE~\cite{open-swe} workflow in
Figure~\ref{fig:agent_workflow}. A program-manager agent decomposes a
request into subtasks, developer agents generate code and invoke tools, and
testing agents evaluate the generated code and may trigger another
developer invocation. The execution structure is dynamic: branches,
fan-out, and retries depend on intermediate results, and available
parallelism may become known only during execution. Components are also
heterogeneous, ranging from millisecond-scale tool calls to LLM generations
and long-running sandboxes with different resource requirements. In
addition, LLM invocations can leave behind KV cache state that may be
useful when later workflow stages or retries return to the same model.

\begin{figure}[t]
    \centering
    \includegraphics[width=0.8\linewidth]{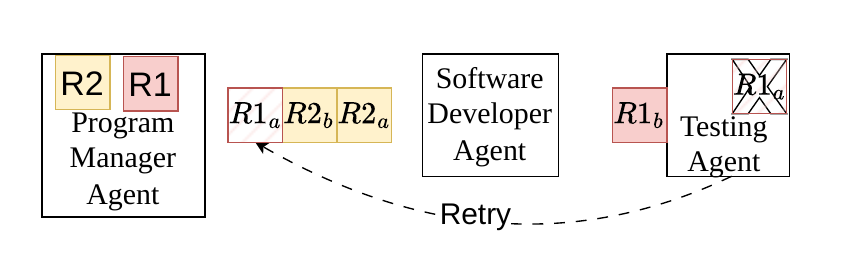}
    \vspace{-10pt}
    \caption{\textbf{Challenges serving agentic application:} 
    Serving concurrent agent workflows without global orchestration. Independent component scheduling leads to head-of-line blocking and wasted recomputation, demonstrating the need to account for workflow progress as well as cluster-wide resource information. 
    }
    \label{fig:agent_challenges}
    \vspace{-15pt}
\end{figure}

These properties make workflow-level context relevant even to decisions made at
individual components. Figure~\ref{fig:agent_challenges} illustrates this
with two requests, $R_1$ and $R_2$, that share a developer agent. Suppose
$R_1$ has progressed through development and testing, fails a test, and
returns to the developer. A scheduler using only the developer's local queue
may place this retry behind newly arrived work from $R_2$, even though
completing the retry may finish $R_1$. Similarly, if $R_1$ is blocked on
another component, deciding whether to retain or offload its KV cache
depends on when the workflow is likely to return to the corresponding LLM.
Such decisions depend on both the request's workflow progress and resource
conditions  in the cluster.

We therefore argue that agent serving requires two complementary forms of
runtime information. A \emph{workflow-level view} must capture the agent and
tool invocations associated with each request, their dependencies, and their
execution progress, and must evolve as new invocations and dependencies are
created. A \emph{cluster-wide resource view} must capture run-time queueing,
component load, available compute and memory, and the KV cache placement. Because workflows evolve and operations span timescales from
milliseconds to minutes, the runtime must combine these views using
low-overhead local decisions and periodic global control that can revise
placement, prioritization, resource allocation, and KV cache management 
decisions as workflow execution and cluster conditions change.

Existing systems provide only parts of these capabilities. Frameworks such
as CrewAI~\cite{crew}, LangGraph~\cite{langgraph}, and
AutoGen~\cite{wu2024autogen} simplify the construction and execution of
agent workflows, but provide limited cluster-wide scheduling and resource
control. Systems such as Autellix and KVFlow~\cite{luo2025autellix,
pan2025kvflow} use application context to improve LLM inference, but do not
coordinate execution across the heterogeneous agents and tools that
determine end-to-end workflow performance. Ayo~\cite{tan2025towards}
performs end-to-end optimization when the execution graph is known in
advance, which limits its applicability to workflows whose branches and
retries are determined during execution. Generic distributed runtimes such
as Ray~\cite{moritz2018ray} support dynamic task execution, but expose
limited workflow context for revisiting scheduling and placement decisions.
Existing systems therefore do not simultaneously provide dynamic workflow
visibility, cluster-wide control, and the programming flexibility expected
by agent developers.

We introduce \system, an agent-serving system designed to meet these
requirements. \system's first key idea is to expose dynamic workflow structure
without requiring developers to declare a graph. \system automatically
generates lightweight stubs that interpose on calls between agents and tools
while preserving their Python interfaces. 
Each invocation creates a \emph{future} carrying workflow context that is augmented with dependency, consumer, and placement information as execution proceeds.
% Each invocation creates a
% \emph{future} carrying workflow metadata, including its request identity,
% dependencies, creator, consumers, and assigned executor. 
As invocations are
created and consumed, these futures capture the dependency structure and
progress of the request. They also decouple creation from execution, allowing
the runtime to revise routing, priority, and executor choices after an
invocation has been issued.

\system's second key idea is a two-level control architecture that systematically combines the above
workflow information with cluster-wide state. Component-level
controllers colocated with agents and tools handle latency-sensitive events
such as future creation, readiness, and dispatch. A global controller runs
off the critical path, aggregates workflow and resource telemetry, and
periodically revisits routing, prioritization, migration, and resource
allocation. 

Thirdly, \system also uses workflow progress to manage LLM inference
state: dependency progress and workflow liveness guide KV-cache retention,
offloading, prefetching, and reclamation. Thus, the workflow information is
used both to schedule computation and to also to inform management of the inference state needed by that computation.

Together, \system's mechanisms support a \emph{workflow-aware orchestration
policy} for end-to-end execution under a fixed resource budget. Our default
policy combines late-binding load balancing, dynamic resource reallocation,
workflow-aware prioritization, and KV-cache management using workflow
progress and cluster state. Importantly, the underlying same control interface also allows operators
to modify the policy without changing workflow code.
In summary, \system turns dynamically generated agent invocations and their state into revisable runtime objects, enabling global (workflow- and cluster-level) policies without placing global coordination on the critical path.

We implement \system in approximately 9,600 lines of Python and evaluate it
on three representative agentic workflows. Relative to existing agent
frameworks, \system improves average latency by up to $3.38\times$ and
reduces P95--P99 latency by 34--74\%, while sustaining offered loads at
which several baselines overload. Ablations show that both workflow-aware
futures and cluster-wide control contribute substantially to these gains,
while the two-level architecture keeps critical-path scheduling overhead
low at scale.

\section{Background and Motivation}
\label{sec:background}

%\aditya{Reframed the section opener to flow directly from the intro. The goal here is no longer to restate the main thesis of the paper, but to make the workload structure concrete, extract the systems-relevant properties, and set up the design without redundancy.}
Using an example, we highlight agentic workflow structure, the underlying challenges, and the ideas behind our design.

\subsection{Agentic Workflows}
\label{subsec:agent_background}

%\aditya{Rewrote the opening around how much workflow structure is known ahead of time. This gives the router-vs.-planning distinction a concrete systems purpose, instead of making it read like an arbitrary taxonomy. It also lets us identify the running example precisely and connect it back to the control challenges from the intro.}
% Agentic workflows differ not only in what components they invoke, but also in how much of their execution structure is known in advance. 
% At one end are \emph{semantic-router workflows}, which make an upfront routing decision and then execute one of several mostly fixed downstream pipelines. At the other end are \emph{planning workflows} in which an LLM or heuristic repeatedly decides what to do next based on intermediate results and external state. The workflow type generally determines how much of the execution graph the serving system can anticipate ahead of time. 
We start with a running example of Open-SWE workflow. %dynamic end of this spectrum.

\noindent\underline{\emph{Example:}}
%\aditya{Made the example explicitly a planning workflow and removed vague phrases like ``this paradigm.'' The walkthrough is now more compact and more clearly tied to execution structure, state reuse, and retries, which are the properties that matter for serving.}
Figure~\ref{fig:agent_workflow} shows a planning-style software-engineering workflow adapted from OpenSWE~\cite{open-swe}. The program-manager agent decomposes a request into subtasks (Steps~\Circled{1}--\Circled{2b}), which software-engineer agents implement using tools such as documentation search and web search (Steps~\Circled{3}--\Circled{4}). A testing agent retrieves artifacts, executes tests, and returns the results (Steps~\Circled{5}--\Circled{8}). Failed implementations trigger a corrective loop (Step~\Circled{9}) that may reuse artifacts and inference state from prior attempts; successful implementations are returned to the user. This workflow exhibits three properties that complicate serving: data-dependent execution creates dynamic branches and retries; KV-cache state may outlive an invocation and be reused when execution returns to the same LLM-backed agent; and heterogeneous components, including LLMs, search tools, and test environments, have widely different latency and resource profiles.

\subsection{Observations}

%\aditya{Rewrote the lead-in so it no longer repeats the intro or Section~\ref{subsec:agent_background}. The purpose here is now explicit: distill the example into a small set of systems observations that directly motivate the design.}
% The example above points to three key observations. %that guide the design of \system.
% Before making additional observations we would like to highlight that, Figure~\ref{fig:agent_challenges} highlights that decisions made at an individual component can depend on where a request is in its workflow and on resource conditions elsewhere in the cluster. Thus, effective orchestration requires combining workflow-level execution context with a cluster-wide view of queues and resources
% Now we highlight other important observations from the workflow. 
Figure 2 highlights that decisions at an individual component depend on both workflow progress and resource conditions elsewhere in the cluster. Against this backdrop, 
Next, we make three observations that shape \system's design.

\noindent\textbf{Implicit parallelism.}
%\aditya{Sharpened this observation to focus on what is novel from a systems perspective: agent workflows often contain exploitable parallelism, but current frameworks make developers surface it manually with workflow-specific concurrency code. The takeaway is now about extracting concurrency from workflow structure, not vaguely about ``coordination.''}
Agentic workflows often contain substantial parallelism that is implicit in the program but invisible to the runtime. In Figure~\ref{fig:agent_workflow}, 
tasks generated by the programming agent are independent and can proceed concurrently. More generally, a planning workflow may spawn multiple subtasks, tool invocations, or validation stages whose dependencies become clear only as execution unfolds. Current frameworks require developers to expose such concurrency explicitly using async calls, threads, or workflow-specific orchestration logic. This is cumbersome and brittle: the developer needs to reason about dependencies manually. 
% that are already present in the workflow semantics.

\noindent\underline{\emph{Takeaway:}}
The runtime should recover dependency structure from workflow execution and use it to schedule independent work concurrently, without requiring developers to encode concurrency explicitly.

%\begin{figure}
%    \centering
%    \includegraphics[width=0.8\linewidth]{figures/token_generated_agent.pdf}
%    \caption{Token generation CDF for two agents in the OpenSWE workflow.}
 %   \label{fig:cdf_resource_utilization}
%\end{figure}

\begin{figure}[t]
    \centering
    \begin{subfigure}[t]{0.47\columnwidth}
        \centering
        \includegraphics[width=\linewidth]{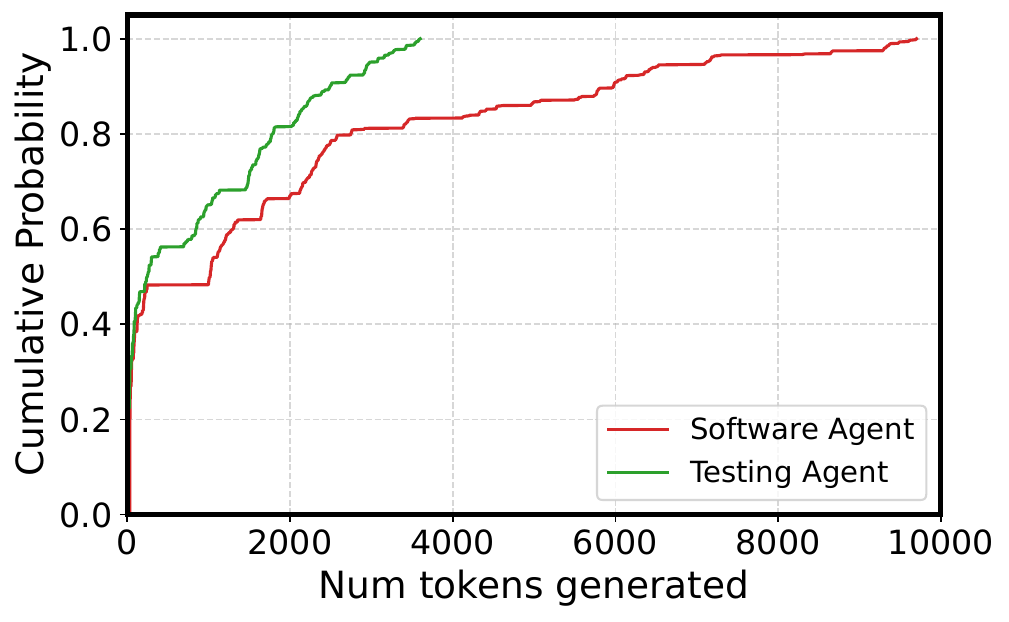}
        \caption{Generation length CDF for two agents in OpenSWE.}
        \label{fig:cdf_resource_utilization}
    \end{subfigure}\quad
    \begin{subfigure}[t]{0.47\columnwidth}
        \centering
        \includegraphics[width=\linewidth]{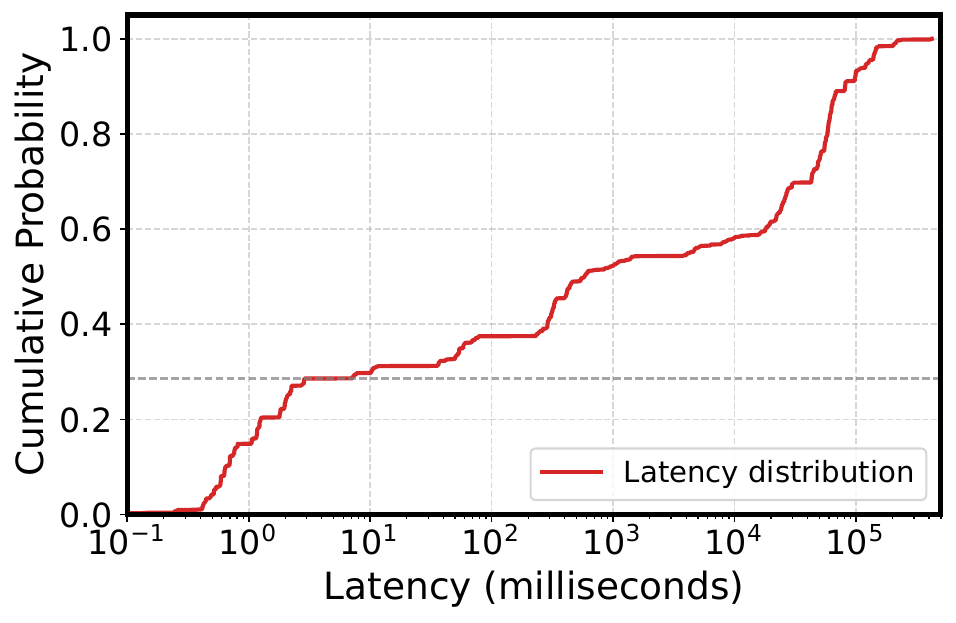}
        \caption{CDF of tool invocation latency OpenSWE (log scale).}
        \label{fig:runtime-tool-cdf}
    \end{subfigure}
    \vspace{-10pt}
\caption{Workload heterogeneity in a workflow.} %\saurabh{icnrease label size}}
    \label{fig:workload_heterogeneity}
\vspace{-15pt}
\end{figure}

\noindent\textbf{Variable execution costs and head-of-line blocking.}
Execution times vary widely across components and invocations. Figure~\ref{fig:cdf_resource_utilization} shows that OpenSWE agents have distinct generation-length distributions, with invocations of the same agent spanning over two orders of magnitude in generated tokens. Similarly, Figure~\ref{fig:runtime-tool-cdf} shows that over 30\% of tool calls complete within 3\,ms, while some exceed 5 minutes. This variability makes static allocation and one-time placement insufficient. Across 1,000 SWE-bench requests scheduled on three testing-agent instances using shortest-queue routing, 24.7\% still wait behind long-running invocations because queue length does not capture remaining execution time. It also necessitates multiple control timescales: global coordination is too expensive for millisecond-scale operations, while local scheduling alone cannot correct persistent imbalance.

% Execution times vary sharply both across components and across invocations
% of the same component. Figure~\ref{fig:cdf_resource_utilization} shows that
% OpenSWE's software-engineer and regression-testing agents have different
% generation-length distributions, while invocations of the same agent span
% more than two orders of magnitude in generated tokens.
% Figure~\ref{fig:runtime-tool-cdf} shows similar variation across tools:
% over 30\% of tool calls complete in under 3\,ms, while the longest-running
% operations take over 5 minutes.
% This variability makes both static resource allocation and one-time request
% placement insufficient. It also causes \emph{head-of-line blocking} even
% when initial placement accounts for queue lengths. We study 1,000
% SWE-bench requests across three testing-agent instances using shortest-queue
% routing. Despite balancing queue lengths at dispatch time, 24.7\% of
% requests still wait behind a long-running invocation because queue length
% does not capture unknown remaining execution time. At the same time, the
% wide range of execution times makes a single control timescale unsuitable:
% cluster-wide control on every millisecond-scale operation would add
% excessive overhead, while a purely local fast path cannot correct persistent
% load imbalance.

\noindent\underline{\emph{Takeaway:}}
Agent serving requires low-overhead local execution for short operations
together with periodic global control for longer-lived work. Runtime
must also adapt resource allocation to observed demand and support
\emph{revisable placement}, allowing queued work to move when execution
progress or queueing conditions invalidate initial decision.

%\begin{figure}
%    \centering
%    \includegraphics[width=0.8\linewidth]{figures/cdf_latency_tool_call.pdf}
%    \caption{CDF of tool invocation latency in the OpenSWE workflow.}
%    \label{fig:runtime-tool-cdf}
%\end{figure}

\noindent\textbf{KV Cache Management.}
Agentic workflows generate substantial implicit execution state (e.g., the KV cache created during LLM inference). Effective orchestration therefore must extend beyond compute scheduling to govern these state lifecycles directly. Take two concurrent requests, $R_1$ and $R_2$. If $R_1$ suspends LLM generation to invoke a long-running external tool, it stops consuming active compute, yet its KV cache remains pinned in GPU HBM, creating a critical tension. If the runtime retains the cache, it may starve $R_2$ of memory and leave compute idle; if it blindly evicts the cache to admit $R_2$, resuming $R_1$ incurs substantial recomputation latency. A workflow-aware control plane resolves this by decoupling state from compute: offload $R_1$'s KV cache to lower memory tiers during the stall and prefetch it back to HBM before execution resumes.

\noindent\underline{\emph{Takeaway}}
Effective agent orchestration must extend beyond mere compute allocation; it requires explicit control over the inference state such as KV caches. Consequently, a robust serving architecture must elevate state management to a first-class control plane primitive.
 
These observations motivate a design that exposes workflow structure to the runtime, supports policy-driven control, and separates fast local decisions from slower global ones.

\subsection{Existing Agent Serving Systems}
\label{sec:othersystems}
%\aditya{Rewrote the transition so this sudbsection reads as the final motivating step before the full design. It now ties the limitations of prior systems directly to the observations above: lack of dynamic structure, lack of control, and weak state handling.}
Existing agent frameworks expose a gap between what developers can express and what a serving runtime must control. %Broadly, they fall into two categories.

%Existing agent frameworks illustrate the gap between what developers can express and what a serving runtime must control. Broadly, current systems fall into two categories.

\noindent
\textbf{Specification-focused frameworks.}
%\aditya{Tightened this paragraph to emphasize the systems limitation: these frameworks help build workflows, but provide little runtime support for selective scaling, scheduling, or state-aware management.}
Frameworks such as CrewAI~\cite{crew} and AutoGen~\cite{wu2024autogen} focus on workflow specification. They simplify composition of agents and tools, but leave scheduling, resource management, and policy logic to developers who must encode orchestration in application code or replicate entire workflows to handle load, rather than scaling or rebalancing individual bottleneck components.
%Frameworks such as CrewAI~\cite{crew} and AutoGen~\cite{wu2024autogen} focus primarily on workflow specification. They make it easier to compose agents and tools, but leave resource management, request scheduling, and policy implementation outside the framework. evelopers must embed orchestration logic into application code or replicate entire workflows to handle load, rather than scaling or balancing load across individual bottleneck components and their replicas.

\noindent
\textbf{End-to-end frameworks.}
%\aditya{Kept the two-class organization, but made the critique more structured and less repetitive. The three subpoints now map more clearly to the observations above: dynamic structure, control/scheduling, and state.}
Frameworks such as LangGraph~\cite{langgraph}, Parrot~\cite{lin2024parrot}, and Ayo~\cite{tan2025towards} combine specification and execution support, but remain limited in three ways.

%Frameworks such as LangGraph~\cite{langgraph}, Parrot~\cite{lin2024parrot}, and Ayo~\cite{tan2025towards} attempt to provide both specification and execution support, but they remain limited in three ways.

\noindent\underline{\emph{Specification.}}
Ayo requires static workflow graphs, which is poor fit for planning workflows whose structure emerges at runtime through tool results, retries, and model-driven branching. Parrot captures some LLM dependencies via semantic variables, but does not generalize cleanly to flexible tool use or data-dependent control flow.

%Ayo and LangGraph require workflows to be described as static graphs, which is a poor fit for planning workflows whose structure emerges at runtime through tool results, retries, or model-driven branching. Parrot introduces semantic variables to capture some LLM dependencies, but its model does not generalize cleanly to flexible tool use or data-dependent control flow.

\noindent\underline{\emph{Scheduling and resource management.}}
Ayo and Parrot each optimize specific execution behaviors, but neither exposes the control needed for policy-driven serving. Ayo can pipeline and parallelize execution when the graph is known in advance, but that assumption fails for dynamic workflows. Parrot focuses on batching LLM calls, but cannot account for runtime changes in control flow or cross-agent coordination. LangGraph largely defaults to best-effort FCFS execution and provides no principled interface for prioritization, migration, or cluster-level resource steering.

%Ayo and Parrot each provide point solutions for specific execution optimizations, but neither exposes the level of control needed for policy-driven serving. Ayo can pipeline and parallelize execution when the computation graph is known, but this assumption does not hold for dynamic workflows. Parrot focuses on batching LLM calls, but its strategy is too rigid to account for runtime changes in control flow or cross-agent coordination. LangGraph  defaults to best-effort FCFS execution and provides no principled interface for prioritization, migration, or cluster-level resource steering.

\noindent\underline{\emph{KV Cache Management.}}
Most frameworks\cite{tan2025towards, lin2024parrot, luo2025autellix} provide only lightweight request-scoped state. Developers remain responsible for longer-lived session state, including its placement, lifetime, and consistency across retries and interactions. This makes KV cache challenging as LLM serving engine has no context about the workflow requirements. 
% stateful workflows harder to serve correctly and efficiently, especially when requests may move across resources over time.

%Most frameworks provide only lightweight request-scoped state. Developers remain responsible for managing longer-lived session state, including its placement, lifetime, and consistency across retries and interactions. This makes stateful workflows harder to serve correctly and efficiently, especially when requests may need to move across resources over time. %\aditya{this seems a little weak. if we have eval then we can add a fwd pointer}

%\aditya{Added a concluding sentence that explicitly connects the prior-art discussion back to \system's design space. This avoids ending the subsection as a disconnected literature summary.}
%Taken together, these limitations motivate a serving system that can observe dynamic workflow structure, manage state explicitly, and expose programmable control over execution.

\subsection{\system's Key Ideas}

%\aditya{Rewrote the opening so it complements rather than repeats the intro. Instead of re-describing the components in broad terms, it states the design goal crisply: preserve Python programmability while giving the runtime the control surface implied by the observations above.}
\system is built to preserve the flexibility of ordinary Python workflows while giving the runtime enough structure and control to schedule, place, and manage them effectively.

\textbf{Decoupling execution from control via semantic futures.} (\S\ref{sec:system})
 \system decouples \emph{when} workflow code issues an invocation from \emph{when and where} the runtime executes it. \system does this by automatically interposing on agent and tool calls and materializing each invocation as a future enriched with workflow-aware metadata. This turns a dynamic Python workflow into an explicit runtime object: dependencies become visible as they emerge, execution context persists across stages, and placement decisions can be deferred until runtime. %Section~\ref{sec:system} describes how stub generation and semantic futures realize this interface.

\noindent
\textbf{Control split by timescale.} (\S\ref{sec:control}) 
\system's control plane design notes that agent serving should not treat all runtime events uniformly. Short operations require a very low-overhead fast path, while long-lived work benefits from cluster-wide reconciliation. \system therefore separates local event-driven control from periodic global control: component-level controllers react immediately to critical-path events, while a global controller revisits scheduling, routing, and KV cache placement decisions using a broader view of workload and resource conditions. %Section~\ref{sec:control} details this architecture.

\noindent
\textbf{KV cache as a controllable object.} (\S\ref{sec:state_management})
\system elevates KV caches into globally controllable runtime objects, allowing the control plane to preserve session semantics while adapting cache placement, reuse, migration, and offloading to evolving workflow and resource conditions.

\begin{figure}
    \centering
    \includegraphics[width=1.0\linewidth]{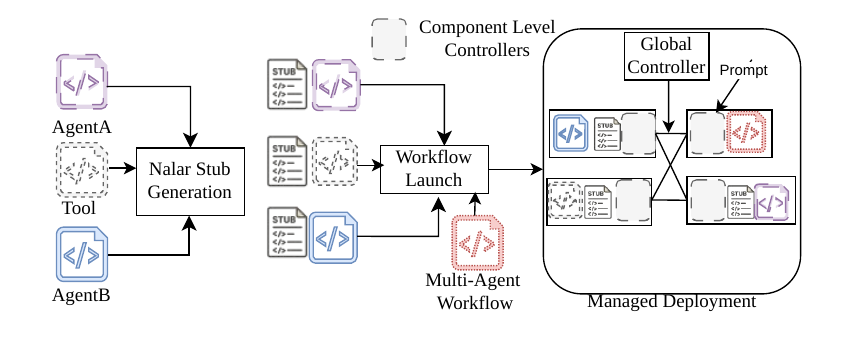}
    \caption{\textbf{\system Overview:}
    \system takes user-specified files and generates stubs (\S~\ref{sec:programming-workflow}, \S~\ref{sec:stub_generator}) that replace original function calls with controllable hooks to generate futures (\S~\ref{subsec:futures}). These stubs connect the user program to the runtime controllers. At deployment, \system launches and manages the runtime (\S~\ref{sec:control}), where component-level controllers and the global controller coordinate to enforce scheduling, routing, and resource policies while managing KV caches.}
    \label{fig:ventis-overview}
    % \vspace{-15pt}
\end{figure}

% By making KV cache state as a controllable object, \system can preserve session semantics while controlling placement, reuse, migration, and offloading as execution paths and resource conditions evolve. %Section~\ref{sec:state_management} describes the resulting managed state layer.
\noindent
Figure~\ref{fig:ventis-overview} provides a full overview of \system's design.

\section{Programming Model}
\label{sec:system}

\system enables developers to express agentic workflows in ordinary Python while exposing enough structure for the runtime to observe dependencies, attach execution context, and control execution. At a high level, \system introduces two mechanisms. First, it generates \emph{stubs} that interpose on cross-component calls without changing the programmer-visible interface. Second, those stub invocations return \emph{futures}, which serve as the runtime objects through which \system tracks dependencies, context, and execution state. We first describe what developers write (\S\ref{sec:programming-workflow}), and then explain how stubs (\S\ref{sec:stub_generator}) and futures (\S\ref{subsec:futures}) realize this model. 

%\system enables developers to express agentic workflows in ordinary Python while exposing enough structure for the runtime to observe dependencies, attach execution context, and control execution. This section describes how the \system interface achieves this in three steps. First, we show how developers specify agents, tools, and workflows. Second, we describe the stub generator that converts ordinary-looking cross-component calls into runtime-visible operations. Third, we describe the future abstraction that NALAR uses to represent and manage those operations.

\subsection{Specifying Agent Workflows}
\label{sec:programming-workflow}
\label{sec:programming}

\system treats agents and tools as standard Python callables and supports dynamic control flow, including nested invocations, dynamic branching, retries, and synchronous execution. The only structural separation is between \emph{component specification}, which defines the implementation of an agent or tool, and \emph{workflow specification}, which defines how a request traverses those components.

\begin{figure}[t]
    \centering
    % \begin{subfigure}{0.45\textwidth}       
        \resizebox{1.0\linewidth}{!}{
\begin{lstlisting}[style=customc2]
import vllm_shared
import technical_documentation as documentation
import testing_agent as tester
class DeveloperAgent:
    def __init__(self):
        self.role = "You are a software developer ..."

    # Tries to generate working code in one-shot
    def implement_and_test(self, task):
        docs = documentation.get(task) 
        generated_code = vllm_shared.execute(self.role + task + docs)
        test_result = tester.unit_test(generated_code)
        return test_result, generated_code
\end{lstlisting}}
        \vspace{-10pt}
        \caption{{\textbf{Example Agent}: A \textit{software developer} agent definition. It calls a documentation lookup tool, a shared inference engine and another testing agent. These calls look like calls to local objects.}}
        % \caption{\small{Building a agent using \system: The above agent uses two tools, a shared vLLM tool and calls another testing agent. It also uses \system's  memory abstraction to store memory.}}
        \label{fig:programming-developer}
    % \end{subfigure}
    \vspace{-10pt}
\end{figure}
\begin{figure}[t]
    % \begin{subfigure}{0.45\textwidth}       
        \resizebox{1.0\linewidth}{!}{
\begin{lstlisting}[style=customc2]
import nalar
import planner_agent as planner
import developer_agent as developer

# Optional hints to guide the runtime
planner.init(preemptable=True, max_resources={"GPU":2, "CPU":1})
developer.init(batchable=True, max_resources={"GPU":4, "CPU":2})

def main(prompt, max_retries):
    # 1. Planner decomposes the request into subtasks.
    subtasks = planner.plan(prompt)
    subtask_done_list = [False] * len(subtasks)
    generated_codes = [None] * len(subtasks)

    # 2. Assign each subtask to a different developer with relevant docs
    # Each developer immediately returns a future
    # which eventually resolves to "Pass" / "Fail".
    futures = [developer.implement_and_test(subtask) 
                        for subtask in subtasks]
    
    # 3. Developer might fail to generate to working code for its subtasks.
    # So the driver implements fine-grained retry logic.
    retries = 0
    while (False in subtask_done_list):
        if retries > max_retries:
            raise Exception(f"Failed to implement {prompt}")

        for i, future in enumerate(futures):
            if not future.available():
                continue
            test_result, generated_code = future.value()
            if test_result == "Pass":
                subtask_done_list[i] = True
                generated_codes[i] = generated_code
            else:
                futures[i] = developer.implement_and_test(subtasks[i])
                retries += 1
    
    # 4. Merge code from each subtask and return it
    return concat(generated_codes)

if __name__ == "__main__":
    deployment = nalar.deploy()
    deployment.main("Enable OAuth login for the website", max_retries=3)
\end{lstlisting}}
        % \vspace{-10pt}
        \caption{{\textbf{Three-agent workflow:} The planner agent decomposes a natural-language coding request into subtasks. Each subtask is sent to a Developer agent from Figure~\ref{fig:programming-developer}, which returns a future indicating test success or failure. The program creates and consumes these futures, retrying failing subtasks. Unless the user explicitly desires(Line 29) they don't need to interact with futures. }}
        \label{fig:programming-driver}
    % \end{subfigure}
    %\caption{\small{A example three-agent coding workflow in \system.}}
    \label{fig:programming-example}
    \vspace{-20pt}
\end{figure}
% \begin{figure}
%     \centering
%     \includegraphics[width=1\linewidth]{figures/system_workflow-compressed.pdf}
%     \caption{\small\textbf{Schematic of Agentic Application:} Each application needs access to different application for 4 \aditya{what on earth is this? drop?}}
%     \label{fig:agent_userflow}
% \end{figure}

\noindent\underline{\emph{Agent \& tool specification.}}
In \system, agents and tools are written as ordinary Python classes and functions. An agent may call tools, invoke other agents, or trigger a nested workflow. \system imposes no fixed execution topology and doesn't require the developer to describe the workflow as a static graph.
Figure~\ref{fig:programming-developer} shows this style using part of the software-engineering workflow from Section~\ref{subsec:agent_background}. The developer agent retrieves documentation, generates code, and invokes a tester agent. From the programmer's perspective, these operations are expressed as standard method calls on local objects. Communication, placement, and orchestration are not encoded in the workflow logic itself.

\noindent\underline{\emph{Workflow specification.}}
A multi-agent workflow is written as a \emph{driver} program that ties together agents and tools and defines how a user request is processed. Figure~\ref{fig:programming-driver} shows the driver for our three-agent example. In Lines~2--3, the programmer imports agents as if they were local modules. Lines~5--7 provide optional runtime directives, discussed later in Section~\ref{subsec:hints}.

The \texttt{main} function illustrates how \system preserves ordinary control flow while interposing on execution. In Part~\#1, the driver invokes the planner agent to decompose the user request into subtasks (Line 11). This invocation returns a \emph{future}, a runtime handle for the pending result, rather than a concrete value.  %This invocation returns a \emph{future} rather than a concrete value. 
The program blocks only when it needs the number of subtasks in the next line (Line 12), since that information is unavailable until planning completes. In Part~\#2, each subtask is passed to a developer agent. These invocations also return futures (Line 18), allowing the subtasks to proceed without serial blocking. In Part~\#3, the driver checks whether individual futures have resolved and inspects their returned values (Line 29). If a developer invocation returns a failed test result, the driver reinvokes that subtask.

The key point is that the same programming model applies throughout. A top-level driver, an agent, or a nested workflow can all invoke other components using the same ordinary Python interface.
The remaining question is how these ordinary-looking calls are converted into stub invocations that return runtime-managed futures without changing the workflow code that developers write. We address this next.

%A central question remains: {\em how are ordinary function calls converted into runtime-managed operations without changing the workflow code that developers write?} \system addresses this through its stub generator.

% The stub generator performs two transformations. 
%inserts an interception boundary between workflow code and component execution.
% Internally, the generator analyzes the Python source to recover callable interfaces and combines that information with the specification to resolve ambiguities arising from dynamically typed code. This step ensures that the runtime exposes the intended methods and packages their arguments consistently, without requiring developers to rewrite component code around a framework-specific API.
\subsection{Stub Generator}
\label{sec:stub_generator}

Ordinary Python calls do not provide a stable control point for distributed orchestration. A local function call cannot, without explicit interposition, be deferred, remotely scheduled, associated with workflow context, or represented as runtime-managed work. \system addresses this limitation through a \emph{stub generator}, which interposes between workflow code and the execution of agent and tool components. Before deployment, the developer provides the Python implementation of each agent or tool. The stub generator parses the source into an abstract syntax tree (AST) and recovers its statically declared structure, including top-level functions, classes, methods, and argument lists. From this information, \system identifies the methods exported for remote invocation and generates two artifacts.
\system today chooses to export all public methods.

First, \system synthesizes an importable client-side module whose exported methods preserve the original component's import path, method names, argument structure, and ordinary call syntax. Invoking a generated method does not execute the component locally. Instead, the stub immediately creates and returns a future representing the pending invocation. At creation time, the stub records the target component and method, invocation arguments, creator, session identifier, and execution context. The runtime subsequently augments this metadata with the assigned executor, dependencies, consumers, readiness state, and placement information.
Second, \system generates a server-side wrapper that exposes each exported method through a standard gRPC service interface. The wrapper is packaged with the original component implementation and its dependencies to form a deployable runtime component. Calls within a component continue to execute as ordinary Python calls; only invocations across  component boundaries pass through generated stubs.

The stub generator therefore serves two purposes. It preserves the programmer-visible interface of ordinary Python components, which keeps workflow development simple. At the same time, it generates precise interception points for every cross-component invocation. That interception point is what allows the runtime to create futures, recover dependency structure, propagate execution context, and later apply policies for routing, prioritization, migration, and KV cache management.
We discuss limitation of Stub Generator in Appendix\S\ref{sec:stub-limitation}.

% We consider these restrictions largely benign. Developers typically rely on such advanced language features to bypass local execution bottlenecks; because \system intrinsically provisions scalable, distributed concurrency under the hood, the need for these localized, manual performance workarounds is mostly mitigated.

\subsection{Futures}
\label{subsec:futures}

Futures are a standard abstraction in distributed systems, with a lineage spanning early programming models~\cite{hewitt1977actors, liskov1988distributed, chatterjee1989futures, bloom2009thorn} and modern runtimes such as CIEL~\cite{murray2011ciel}, Dask~\cite{peters2023parallel}, and Ray~\cite{moritz2018ray}. 
A future represents the eventual result of an asynchronous computation. In \system, futures additionally serve as revisable orchestration objects carrying workflow and execution metadata.

% A future is a proxy for the eventual result of an asynchronous computation. By returning a future immediately upon task dispatch, the runtime provides a non-blocking handle through which downstream computation can proceed, synchronize, or explicitly block for resolution.
% In \system, futures are the core runtime objects through which the system observes, reasons about, and manages workflow execution.

\noindent
\textbf{Futures as the orchestration primitive.}
To build and manage a distributed computation graph without changing developer code, \system requires an abstraction that can represent pending work, defer synchronization, and expose dependencies as execution proceeds over time. Native concurrency mechanisms such as \texttt{asyncio} or local threading can express asynchronous execution, but they force the developer to manage synchronization and control flow explicitly. By pairing futures with the stub generator, \system can intercept cross-component calls transparently, represent each invocation as a first-class runtime object, and expose as much parallelism as possible without requiring manual annotations from the workflow programmer.

% \noindent\textbf{Future }
\noindent
\textbf{Futures Methods.}
By default, \system resolves futures implicitly: evaluating a future blocks until the underlying computation completes. However, some workflows benefit from explicit lifecycle control, such as the non-blocking retry logic in Figure~\ref{fig:programming-example} (Line~29). To support these cases, \system futures expose lightweight methods: (i) \texttt{available()}, a non-blocking readiness check, and (ii) \texttt{value(timeout=t)}, which blocks for up to \emph{t} s before returning the result or raising a \texttt{TimeoutException}.

\begin{table}[t]
\caption{\system's future metadata. }
\vspace{-10pt}
\label{tab:future-metadata}
\resizebox{1.0\linewidth}{!}{
\begin{tabular}{@{}lll@{}}
\toprule
Metadata & Structure & Description \\ \midrule
dependencies & list(agentA:ip, \ldots) & List of dependencies required to compute the future's output \\ \midrule
creator & agentName:ip & Component that created the future \\ \midrule
executor & agentName:ip & Component instance currently assigned to execute the future \\ \midrule
consumers & list(future-id, \ldots) & Components waiting to consume the future's value \\ \midrule
session-id & str:UUID & The user-request associated with the future \\ 
\bottomrule
\end{tabular}
}
\vspace{-15pt}
\end{table}

\noindent
\textbf{Metadata.}
Unlike futures in generic execution frameworks, futures in \system carry workflow-aware metadata. Each future stores its dependencies, creator, current executor, and consumers (Table~\ref{tab:future-metadata}). This information is sufficient for the runtime to route work locally, propagate readiness to downstream consumers, and revise execution decisions later through mechanisms such as migration. The point of the metadata is not merely to encode topological order, but to make each invocation a semantically meaningful control object.

\begin{figure}[t]
    \centering
    \includegraphics[width=1\linewidth]{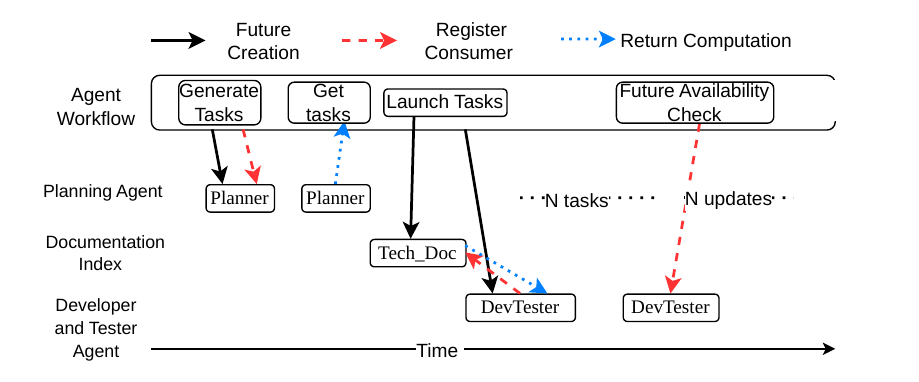}
    \vspace{-15pt}
    \caption{\textbf{Future Generation Timeline:} For the agent workflow depicted in Figure~\ref{fig:programming-example} we depict a timeline for future generation and how their consumers are updated and their values realized in \system.}
    \label{fig:agent_futures_generation}
    \vspace{-15pt}
\end{figure}

\noindent
\textbf{Generation and materialization.}
Figure~\ref{fig:agent_futures_generation} illustrates future creation and consumption in the workflow from Figure~\ref{fig:programming-driver}. Future creation and consumer registration are non-blocking; accessing a future's value blocks until it resolves. For example, the planner invocation creates the \texttt{subtasks} future, which materializes when \texttt{len(subtasks)} is evaluated at Line~12. The resolved subtasks then generate developer futures that materialize when inspected at Line~29. By recording these producer-consumer relationships as they emerge, \system reconstructs the dynamic dependency graph without requiring a statically declared DAG.

% Figure~\ref{fig:agent_futures_generation} illustrates how futures are created and consumed in the workflow from Figure~\ref{fig:programming-driver}. There are three basic operations.

% \noindent
% \emph{Op 1. Future creation.} Creating a future is non-blocking.

% \noindent
% \emph{Op 2. Register consumer.} When an agent or driver depends on a future, it is registered as a consumer -  also non-blocking./

% \noindent
% \emph{Op 3. Return.} Accessing a future's value is blocking.

% When the driver calls the planning agent, a future named \texttt{subtasks} is created. When the driver subsequently evaluates \texttt{len(subtasks)} at Line~12, the future must materialize. At that point, the driver's component controller registers with the planner's component controller as a consumer of that future. Once the \texttt{subtasks} future resolves, the driver receives the subtask list and dispatches each subtask to a developer agent. Each such dispatch creates another future. Later, when the driver inspects the value of a future at Line~29, the same process repeats: the runtime registers the dependency and blocks only when the value is needed.

% This mechanism lets \system recover the evolving dependency graph directly from execution, rather than requiring the workflow to be declared as a static DAG in advance.

% \vspace{0.1em}
\noindent

{\bf Difference from Ray~\cite{moritz2018ray} futures.} 
Although Ray~\cite{moritz2018ray} also uses futures, its \texttt{ObjectRef}s primarily capture task outputs and data dependencies, not the session, workflow progress, downstream consumers, or inference state associated with an invocation. This distinction matters when a request suspends LLM execution on an external tool while its KV cache remains in GPU memory: deciding whether to retain, offload, or prefetch the cache requires knowing whether and when the workflow may return to that LLM. \system embeds this context in futures and combines it with runtime telemetry, enabling workflow-aware KV-cache management, prioritization, late-binding placement, and migration. Implementing equivalent policies in Ray would require explicit propagation of workflow metadata, cross-task state tracking, and custom integration with both the scheduler and inference runtime. Requiring a significant redesign.

\section{\system's Control Architecture}
\label{sec:control}

\S\ref{sec:background} showed that static orchestration is insufficient for agent serving. Efficient execution instead requires a dedicated control plane. 

\subsection{Key Insight \& Design} 

As \S\ref{sec:background} showed, agentic workflows combine two properties that make control difficult: execution times span from sub-milliseconds to tens of minutes, and the execution graph continues to evolve as agents invoke tools, produce intermediate state, and generate new downstream dependencies. This leads to a key design constraint for the control plane. Applying cluster-wide scheduling logic to every short operation would add overhead with little benefit, but 
 % but restricting coordination
% to initial placement is also insufficient because task durations
relying exclusively on static, one-shot placement is inadequate, as task durations are not known \emph{a priori} and good placement decisions can change as workflow structure and cluster conditions evolve. Efficient control, therefore, requires a fast path for short operations together with a mechanism that can revisit and revise decisions for longer-lived work. 
%\aditya{rewrote to clarify, but old text in comments below} \saurabh{made minor change}
%As discussed in \S\ref{sec:background}, subtasks generated by a single agentic workflow exhibit execution latencies spanning from sub-milliseconds to tens of minutes. Our critical insight is that computing globally optimal schedules for every sub-millisecond task introduces prohibitive overhead for very limited benefits; therefore, rigorous scheduling/orchestration should strictly target long-running operations, with policies for short-running tasks to be computed off the critical path. However, this insight alone is insufficient because task durations are rarely known \emph{a priori}. Furthermore, because agentic workflows are highly dynamic, the definition of an "optimal" placement actively evolves as new downstream consumers emerge and cluster utilization shifts. 
Existing distributed frameworks, however, rely on one-shot, event-driven scheduling. In systems like Ray, a task is scheduled exactly once upon creation; this significantly simplifies control logic, as task placement is final and the resulting futures are immutable. \system introduces a different scheduling paradigm for serving agents.
% \aditya{i don't think we can say inspired by SDN as SDN is very broad. moreover it applies to the overall use of control plane, and not the two level stuff} 
We argue that governing agentic workflows requires decoupling control into two distinct mechanisms: \emph{reactive} event-driven placement and \emph{proactive} periodic reconciliation. 
In \system, local controllers handle the event-driven fast path, reacting to future creation by greedily scheduling tasks locally to minimize latency. Concurrently, the global controller executes out-of-band periodic reconciliation logic that analyzes cluster-wide telemetry, revisits prior placement decisions, and dynamically issues directives to migrate, change scheduling rules, or adjust execution priorities.

Our architecture achieves a natural \emph{temporal stratification} of workloads. Transient, millisecond-scale tasks complete their execution via the fast path (without requiring global coordination) and naturally expire before the global controller ever observes them. Conversely, long-running agentic tasks persist across global control epochs, automatically subjecting them to cluster-aware optimization. This design eliminates need for developers to explicitly profile task durations or manually segregate fast and slow execution paths. 

\subsection{Control Components}

%\system as discussed previously, adopts a  {\em two-level control} design that cleanly {\em separates periodic policy computation from event-driven enforcement}.  Shown in Figure~\ref{fig:system-architecture}, the global controller maintains a logically central workflow and system view. It periodically installs scheduling and routing policies at component-level controllers, which apply them immediately as events occur. A node store mediates information flow between the two levels.

\begin{figure}[t]
    \centering
    \includegraphics[width=1\linewidth]{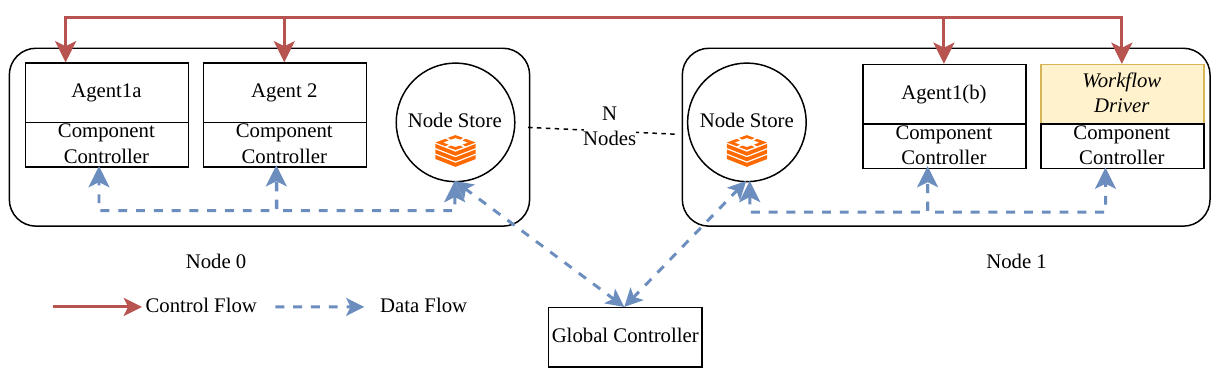}
    \vspace{-10pt}
    \caption{\textbf{\system's two-level control.} Each component has an associated controller with it. Each node has a local node store. The global controller communicates with each agent and workflow driver, through the node store.}
    \label{fig:system-architecture}
    \vspace{-15pt}
\end{figure}

\system's control plane has three components (Figure~\ref{fig:system-architecture}).
% component-level controllers, a global controller, and a node store that mediates information between them. %We describe each in turn.

\noindent{\bf {Component-Level Controllers.}}
When an agent or tool is launched, \system instantiates a component-level controller alongside it. These controllers serve three roles. First, they schedule work locally under policies installed by the global controller, deciding which futures to execute on the associated agent or tool and when. They also maintain and update future metadata needed for correct value propagation and for revising execution decisions such as migration. Second, they serve as the interface between the programming model and the runtime. The auto-generated stubs (\S\ref{sec:programming-workflow}) route each agent and tool invocation through the component-level controller rather than invoking the user-defined implementation directly. This allows \system to intercept invocations, create futures, and coordinate execution. 
For the instance running LLMs, local controllers also manage KV cache. Third, they collect serving-time metrics, including queue lengths, per-request latencies, and local resource usage, which inform the global controller's periodic computations.

\noindent{\bf{Global Controller.}}
For each workflow deployment, \system runs a global controller that implements
the orchestration policy. Running periodically, the global controller
aggregates workflow metadata and resource telemetry from component-level
controllers through the node stores, maintains a global view of the workflow
and the resources assigned to its deployment, computes decisions related to
request routing, prioritization, and resource allocation, and pushes these
decisions to component-level controllers.
The global controller runs a single-threaded, push-based policy loop. The
single-threaded design provides a single authoritative decision stream and
avoids conflicting updates. The push-based design keeps global policy
computation off the critical path.

%The single-threaded design ensures a single decision-maker and a single authoritative update stream, simplifying implementation. The push-based model keeps the global controller off the critical path. 

%. All routing, scheduling, and resource management policies are implemented at this layer; the actions of those policies are then pushed to 

\noindent{\bf{Node Store.}} 
Because the component-level and global controllers operate at different frequencies, \system decouples them through a node-local store, implemented with a local Redis. Component-level controllers asynchronously write future metadata and telemetry to the store and consume policy updates issued by the global controller. This design provides low-latency state sharing without direct controller synchronization or placing global control on the critical path.
% Because the component-level and global controllers operate at different frequencies, \system uses a node-level store to decouple their communication. Each node maintains a local store that serves as both a metadata repository and a telemetry-and-decision broker: component-level controllers push metrics and local observations to it, and the global controller writes policy updates back. In our prototype, this store is implemented with Redis. This design avoids direct controller-to-controller synchronization while providing low-latency access to shared state. Component-level controllers consume policy updates asynchronously, allowing global decisions to propagate without placing the global controller on the critical path. The node store also holds future metadata needed for dependency tracking and execution management.

% We now describe the policy interface and the interaction among local execution and global orchestration to enforce it.

\subsection{Control Interfaces in \system}
\label{sec:policies}
Next, we introduce the control interfaces used by \system.
% To support agentic workloads with different execution structures and service objectives, \system exposes a minimal yet expressive policy interface. Operators write policy programs that inspect cluster telemetry and session state and invoke a compact set of primitives to control routing, prioritization, migration, and resource provisioning.
% for example if we are implementing a quota based priority for a user session, the two threads need to coordinate to share the quota limits. 

\noindent{\bf{Policy Implementation Interface.}} 

\system separates policy specification from enforcement through a compact interface. Operators specify policies using \texttt{route}, \texttt{set\_priority}, and \texttt{migrate} to control request placement, execution order, and future migration, respectively, and \texttt{provision} and \texttt{kill} to modify component capacity. Global controller executes the specified policy by invoking these primitives, while component-level controllers implement the corresponding execution protocols. We observe this small set of primitives is sufficient to implement the workflow- and system-aware policies; we describe \system's default policy and possible extensions in  \S\ref{sec:serving-policies}.

\noindent
{\bf Control Interaction in \system.} 
%\aditya{streamlined the example, but many issues. first, fonts are hard to read. second, it is unclear how parent/dependent/executor are defined and these terms were clearly specified earlier. third, why is the driver the parent? }\saurabh{check, except font fixes}
% \saurabh{it has been talked about briefly in }
% When an agent is created there is some metadata associated with it like 
% Figure~\ref{fig:future_management} illustrates how the global and component-level controllers coordinate to migrate a future. 
% This example shows two points. First,  Second, the global controller operates at the level of policy, while the component-level controllers carry out the protocol needed to realize that policy.
In \system, the global controller selects policy actions, while component-level controllers implement the required protocols. For example, to migrate a queued future \texttt{f2} from \texttt{B:0} to \texttt{B:1}, \texttt{B:0} prevents local execution, redirects or forwards unresolved dependencies, transfers any associated KV-cache state, and updates the future's executor. \texttt{B:1} activates the future only after \texttt{B:0} acknowledges its removal from the source queue. Thus, migration coordinates future metadata, dependencies, and state without placing the global controller on the critical path. Further details appear in Appendix~\ref{sec:controller-interaction}.

\subsection{Control Hints}
\label{subsec:hints}
\begin{table}[t]
\caption{\system's hint interface }
\vspace{-10pt}
\label{tab:hint-agent}
\resizebox{0.95\linewidth}{!}{
\begin{tabular}{@{}lll@{}}
\toprule
Hint           & Values        & Descriptions                                                                                                                                        \\ \midrule
stateful      & Boolean    & \begin{tabular}[c]{@{}l@{}}True indicates for a session successive calls to \\ the agent will be routed to the same instance\end{tabular}           \\ \midrule
batchable      & Boolean   & True indicates that module can accept a batch of request                                                                                            \\ \midrule
reusable & Boolean & \begin{tabular}[c]{@{}l@{}}Resource created during a session can be reused  \\ by subsequent
invocations within the same session. \end{tabular} \\ \midrule
preemptable    & function & \begin{tabular}[c]{@{}l@{}} True indicates that running invocations of the component \\  may be
preempted by the runtime.\end{tabular} \\ \midrule
max\_instances & Integer & \begin{tabular}[c]{@{}l@{}} Indicates the max number of instances \\ to initialize \end{tabular}                                                                           \\ \midrule
min\_instances & Integer & \begin{tabular}[c]{@{}l@{}}Indicates the min number of instances \\ the framework should keep alive       \end{tabular}                                                                        \\ \midrule
max\_resources & Dict          & Max CPU, GPU, and memory resources that may be allocated to the component.                                                                                                   \\ \bottomrule
\end{tabular}
\vspace{-20pt}
}
\end{table}
% \aditya{first say what is the fundamental reason you need hints. otherwise it feels like you pulled it out of nowhere.}

% \aditya{the beginning 1-2 lines can be improved to motivate hints better. Why is the future/state programming model not enough?}
%Futures give the runtime the hooks it needs for control and visibility, but they are only part of the picture. 
Agents often have execution properties that the runtime can exploit for efficiency. For example, if an agent supports batching, a common pattern in ML workloads, \system can coalesce compatible futures and execute them together, as the throughput of LLM output generation greatly benefits from batching ~\cite{agrawal2024taming,kwon2023efficientmemorymanagementlarge}. 
Incorporating such characteristics can enable higher efficiency.
% Similarly, \system can use information To provide these 
To this end, \system  provides a {\em directive} interface.
Table~\ref{tab:hint-agent} lists the supported agent-level directives used by the runtime to guide execution. Most directives are straightforward, but we highlight the \emph{stateful} directive. For agents marked stateful, \system guarantees that all  requests to the agent that are associated with a single user request (session)
are scheduled in order and routed to the same agent instance, ensuring consistent processing. The stateful hint refers to component-local state that \system does not migrate; it is distinct from KV cache state, which is explicitly managed as described in \S\ref{sec:state_management}. We provide details of hint-interface can improve performance in Appendix\ref{sec:impact-hints}.

\subsection{Managing LLM Inference State}
\label{sec:state_management}
As discussed in \S\ref{sec:background}, KV-cache reuse reduces the cost of repeated LLM invocations but increases GPU memory pressure~ \cite{cheng2025lmcache, stojkovic2025dynamollm}. Existing inference engines~\cite{kwon2023efficientmemorymanagementlarge, zheng2024sglang} manage caches within individual model-serving instances, without visibility into the broader workflow. \system instead associates each cache with its originating session and future, while component-level controllers report its location, size, and transfer cost. Using this context, the global controller manages cache placement and lifetime across workflow execution. We extend vLLM~ \cite{kwon2023efficientmemorymanagementlarge} with block-level interfaces for cache inspection, migration, offloading, prefetching, and invalidation. These mechanisms decouple inference-state placement from the worker that created it without changing the LLM interface exposed to agents. Algorithm details are discussed in \S\ref{sec:policies}.

\section{Serving with \system}
% We briefly highlight telemetry collected and the serving policies in \system.
\subsection{Telemetry collected}
\label{sec:telemetry}
Each agent instance is colocated with a local controller that records the telemetry including future metadata, lifecycle timestamps, queue state, and resource utilization. The global controller aggregates this telemetry to reconstruct each session's execution graph and a flattened workflow graph that preserves component visit frequencies and execution costs. From these graphs, it derives per-session and per-component metrics, including dependency-wait, queueing, and execution times, ready-future counts, and available capacity. Metrics provide the workflow and system visibility used by policies in \S\ref{sec:serving-policies}. Table~\ref{tab:telemetry} in the Appendix provides more details.

\subsection{Workflow- and System-Aware Serving Policy}
\label{sec:serving-policies}

% We use \system's control plane to implement a workflow-aware orchestration
% policy that seeks to  minimize end-to-end workflow completion time under the resource budget
% assigned to the workflow deployment. The policy combines four mechanisms operating at
% different levels: late-binding placement balances requests across replicas;
% dynamic resource allocation shifts capacity across workflow components;
% workflow-aware prioritization determines which ready requests execute first;
% and KV cache management coordinates inference-state residency with expected
% workflow progress. `
We use \system's control plane to minimize end-to-end workflow completion time under a fixed resource budget. Its default policy combines late-binding placement, dynamic resource allocation, workflow-aware prioritization, and KV-cache management. The global controller coordinates these decisions, while component-level controllers enforce them locally.
All four mechanisms use the workflow and system telemetry
described in \S\ref{sec:telemetry}, but act on different control decisions.
We describe each below. 

% For all four mechanisms, the global controller computes decisions using
% workflow and system telemetry, while component-level controllers enforce
% those decisions locally.

\noindent\underline{\emph{Late-binding Load Balancing:}}
\system initially distributes requests across replicas while respecting the
hints in Table~\ref{tab:hint-agent}; recursive calls and retries to stateful agents
remain on the appropriate instance. For stateless agents, variable execution
times can produce persistent queue imbalance even when requests were
initially distributed evenly. When the queue-length difference between two
replicas exceeds $\alpha$, \system considers migrating a queued future to the
less-loaded replica. It performs the migration only when the expected
reduction in queueing delay exceeds the cost of moving the future and its
dependencies. \system estimates this cost from the associated data volume and
observed transfer times. Thus, placement remains revisable after a future is
created, allowing \system to correct transient load imbalance using current
queue and dependency information. Algorithm~\ref{alg:late-binding} gives the
complete procedure. For all our experiments we use a fix $\alpha$ of 20\%, i.e., futures are reassigned if queue length difference is greater. 

\noindent\underline{\emph{Dynamic Resource Allocation:}}
Replica-level balancing cannot address a sustained capacity mismatch across
different workflow components. \system uses its workflow and
cluster-wide resource view to identify component bottlenecks and
shift resources toward them, following ideas from prior work on stream
processing~\cite{kalavri2018three}. For each component $c$, the global
controller estimates its effective service capacity $\mu_c$ from the observed
completion rate of its replicas under the current allocation. From observed
workflow executions, it also computes $V_c$, the average number of times
component $c$ is visited per workflow request. We define
$\widehat{X}_c = \mu_c/V_c$ as the component's \emph{workflow-normalized
capacity}, which estimates the end-to-end workflow throughput that component
$c$ can sustain. The component with the smallest $\widehat{X}_c$ is 
the candidate bottleneck.
\system distinguishes true capacity bottleneck from an upstream dependency
stall by observing ready futures. A component with no ready futures
cannot benefit from additional capacity even if it has many blocked futures.
For a persistent bottleneck, \system considers transferring resources
from underutilized components. Reallocations that satisfy
the components' resource and min/max-instance constraints and are feasible
under the deployment's available resources are considered. Among these feasible
reallocations, \system performs a transfer only if it increases the minimum
workflow-normalized capacity without making the donor the new bottleneck.
To avoid oscillations, bottleneck must persist for four
consecutive control intervals. Algorithm~\ref{alg:block} provides the full
procedure.

\noindent\underline{\emph{Workflow-Aware Request Prioritization:}} 
Within each component, \system uses workflow progress to decide which
ready requests execute first. We measure progress as the number of
completed invocations in the request's session, and prioritize requests
with greater progress. This favors in-progress workflows, including cyclic
retries such as a failed test returning to a developer agent, over newly
arrived work. When priorities are equal, requests are served in arrival
order. \system can additionally favor a request whose KV cache is already
resident at the target worker.
% Within each component, \system uses workflow context to decide which ready
% requests should execute first. In particular, it gives higher priority to
% requests that have already made substantial workflow progress, including
% cyclic retries such as a failed test returning to a developer agent. It can
% also favor a request whose KV cache is already resident at the target worker.
% These decisions reduce end-to-end completion time by avoiding unnecessary
% delay for in-progress workflows and, when possible, reusing already-resident
% inference state.

\noindent\underline{\emph{Workflow-Aware KV-Cache Management:}}
Finally, \system uses workflow progress to manage KV-cache residency under
GPU memory pressure. When a workflow is suspended on unresolved dependencies,
the global controller estimates how long the corresponding LLM state is
likely to remain idle. It offloads a KV cache only when the estimated time
until reuse exceeds the combined cost of offloading and restoring it. Similar
to prior work~\cite{cheng2025lmcache, agarwal2026symphony}, cache blocks are
transferred incrementally to host memory.
The same dependency information guides restoration. For each unresolved
dependency, \system estimates remaining execution time using the historical
average for futures executed by the corresponding component. When the
estimated time until the dependencies resolve falls below the cache
restoration time, \system asynchronously prefetches the cache so that
restoration overlaps with the remaining dependency execution. \system also
uses workflow liveness to reclaim state promptly: 
when the session completes
% when a workflow has no
% outstanding futures or possible downstream invocations, 
the global controller
invalidates its KV cache rather than waiting for memory pressure to trigger
an LRU eviction. Our current remaining-time estimator is intentionally simple;
more sophisticated request-aware prediction is orthogonal to the control
mechanisms described here.

% \noindent\textbf{Implementation.}
% The complete orchestration policy requires 141 lines of policy code using the
% control primitives in \S\ref{sec:policies}, with no changes to application
% code or \system's runtime mechanisms. More importantly, the four mechanisms
% rely on the same three capabilities: workflow structure and execution context
% exposed through futures, cluster-wide visibility into workflow and resource
% state, and runtime control over placement, prioritization, migration, resource
% allocation, and KV cache residency.
% \vspace{-1em}
\subsection{Fault Tolerance}
\label{sec:fault-tolerance}
\noindent\textbf{Component-Level Controller Failure.} Each component-level controller periodically writes a heartbeat to its node-local store. After three missed heartbeats, the global controller relaunches the affected controller and component, removes the instance from the routing set, and redirects new stateless invocations and sessions to healthy instances. Active stateful sessions remain bound to their assigned instance, consistent with Table~\ref{tab:hint-agent}, and must wait for recovery or fail. Pending invocations are marked as failed and the failure is propagated to the workflow; transparent recovery of such work is left to future work. 

\noindent\textbf{Global Controller Failure.} Rather than implementing a new election protocol, \system uses etcd~\cite{etcd} for lease-based controller election. A recovery process on each node participates in the election; the elected process launches the global controller and renews its lease. If the lease expires, another process acquires leadership and launches a replacement. The replacement controller reconstructs its view of active futures, queues, resource assignments, and KV-cache locations from component-level metadata. Meanwhile, component-level controllers continue serving requests using their last installed policies. Global policy execution resumes once reconstruction completes.

\section{Evaluation}
% \vspace{-1em}
\label{sec:evaluation}
\noindent
{\bf Implementation.}
We implement \system in roughly 9{,}600 lines of Python, leveraging several existing libraries: (1) gRPC as the communication backend for all inter-component interactions; (2) vLLM for serving LLM models; (3) ChromaDB, a vector search engine used in our workflows;  and (3) Redis as the node-local store to provide transactional support and reduce coordination overhead between controllers.

\noindent\textbf{Workflows.}
We study three workflows. One static, one semantic router, and one planner-based.
\noindent\underline{\emph{Financial Analyst:}} In this workflow~\cite{dong2025large}, specialized agents analyze stocks, bonds, markets, and news, and an analyst agent summarizes their results. Users may issue delayed follow-up queries, creating a human-in-the-loop workflow with repeated access to a shared LLM, reusable KV-cache state, and resource contention. We evaluate it using FinQA~ \cite{chen2021finqa}.
% In this workflow~\cite{dong2025large}, an analyst agent invokes a stock analysis agent, a bond market agent, a market research agent, and a web/news search agent. The aggregated results are summarized for the user, who may issue follow-up queries after long delays, making this a human-in-the-loop workflow. This workflow repeatedly invokes a shared LLM within a session, creating reusable KV state and resource contention. We use the FinQA dataset~\cite{chen2021finqa} for evaluation.
\noindent\underline{\emph{Router-based workflow:}} This workflow follows a common pattern in which a lightweight agent classifies each query and routes it accordingly either to a chat workflow or, for coding tasks, to a dedicated coding agent. We evaluate this workflow using Microsoft Azure LLM traces~\cite{stojkovic2025dynamollm}, which report request volumes for two distinct workflow types.
\noindent\underline{\emph{Software engineering workflow:}} 
This recursive planning workflow mirrors Figure~\ref{fig:agent_workflow} and uses web search and ChromaDB for tool calls and documentation. The planner has a dedicated LLM, while the remaining agents share another LLM(Fig~\ref{fig:agent_workflow}). We evaluate it on SWE-bench~\cite{jimenez2023swe}.

% This planning workflow mirrors the recursive structure shown in Figure~\ref{fig:agent_workflow}, where test failures trigger repeated developer invocations. We integrate web-search tools and use ChromaDB to store documentation. The planner uses a dedicated LLM, while the remaining agents share another LLM, enabling KV-cache reuse across recursive invocations. We evaluate the workflow using SWE-bench~\cite{jimenez2023swe}.

% This is a planning workflow which mirrors the structure in Figure~\ref{fig:agent_workflow}. We integrate tool calling via web-search APIs and store documentation in ChromaDB. As show in Figure, planner agent has its own LLM. While other agents share an LLM. We evaluate this workflow on the SWE-bench dataset~\cite{jimenez2023swe}. Unlike other workflows, this is recursive. 

% \paragraph{Goal of evaluation.}
% In our evaluation we study the following questions - 
% \begin{itemize}
%     \item How \system adaptively tailors resources to the dynamic workflow structure.
%     \item Under varying workload, can \system quickly adapt to evolving load requirements ?
%     \item How does control over state benefits \system ? 
%     \item Eease of iright policies in \system.
%     \item What is the overhead of \system's control ? 
% \end{itemize}

% Both the global and local controllers are implemented as standalone gRPC servers. This enables developers to provide performance differ
% \subsection{Setup}

\noindent\textbf{Baselines.} We compare \system against several baselines. 
% \noindent\underline{\emph{Ayo}}~\cite{tan2025towards} is an end-to-end optimization framework that represents LLM applications as fine-grained task graphs and exploits parallelism and pipelining over a Ray-based execution engine. \noindent\underline{\emph{CrewAI}}~\cite{crew} is a role-based framework for composing agents and tasks into multi-agent applications. \noindent\underline{\emph{AutoGen}}~\cite{wu2024autogen} is an event-driven framework in which agents coordinate through message-based interactions. \noindent\underline{\emph{LangGraph}}~\cite{langgraph} is a graph-based orchestration framework for stateful, long-running agent workflows. \noindent\underline{\emph{Autellix}}~\cite{luo2025autellix} is an LLM-agent serving system that uses program-level context to schedule dependent LLM invocations.
\noindent\underline{\emph{Ayo}}~\cite{tan2025towards} is an end-to-end optimization framework that represents LLM applications as fine-grained task graphs and exploits parallelism and pipelining over a Ray-based execution engine. 
\noindent\underline{\emph{CrewAI}}~\cite{crew} is a role-based framework for composing agents and tasks into multi-agent applications. 
\noindent\underline{\emph{AutoGen}}~\cite{wu2024autogen} is an event-driven framework in which agents coordinate through message-based interactions. 
\noindent\underline{\emph{LangGraph}}~\cite{langgraph} is a graph-based orchestration framework for stateful, long-running agent workflows. \noindent\underline{\emph{Autellix}}~\cite{luo2025autellix} is an LLM-agent serving system that uses program-level context to schedule dependent LLM invocations.
Together, these baselines cover graph-based application optimization, role-based and message-driven agent frameworks, stateful workflow orchestration, and program-aware LLM inference serving.
% We configure all systems with the same models, hardware resources, inference engine settings, and offered workload whenever their interfaces permit.

% \noindent\textbf{Baseline Configuration}
\noindent\textbf{Baseline Configuration.} We made a best-effort attempt to optimize every baseline. Except for Ayo~\cite{tan2025towards}, the evaluated baselines do not support dynamic resource allocation. For each workflow and offered load, we therefore performed a sweep over feasible resource configurations and report the best-performing configuration. All baselines use vLLM (v0.12) with LMCache-enabled offloading, ensuring that performance differences arise from the serving framework rather than the underlying LLM inference engine.
% \noindent\underline{\emph{Ayo}}~\cite{tan2025towards} enables developers to specify agentic applications using a graph-based interface. It enables parallel execution and pipelining of different components in an agent serving pipeline. {\em Ayo uses Ray to build the execution engine. }
% \noindent\underline{\emph{CrewAI}}~\cite{crew} is a popular library to build agents (with over 41K stars on GitHub). It provides a development framework to build and orchestrate agents.
% \noindent\underline{\emph{AutoGen}}~\cite{wu2024autogen} is a popular library by Microsoft (over 52K stars) to support event-driven programming to build agents. 
% \noindent\underline{\emph{Langgraph}}~\cite{langgraph} Langgraph is a popular deployment library 
% to build low-level orchestration framework designed to build stateful, long-running, and production-grade AI agents with over 40k github stars. 
% % specifically designed for  nother (over 52K stars) to support event-driven programming to build agents. 
% \noindent\underline{\emph{Autellix}}~\cite{luo2025autellix} is  an efficient inference serving engine designed specifically for large language model (LLM) agents running as general programs. It primary only manages LLM inference.  

\noindent\textbf{Experimental Setup.}
All experiments use 2 nodes, 
each with 4 NVIDIA A100 GPUs (80GB HBM), 256GB DRAM, and 4TB SSDs connected via a 100Gbps Ethernet link. We use workflow-specific fine-tuned LLaMA-3-8B models. 
\subsection{End-to-End Evaluation}
\label{sec:end-to-end}

\begin{figure*}[t]
    \centering
      \includegraphics[width=0.7\textwidth]{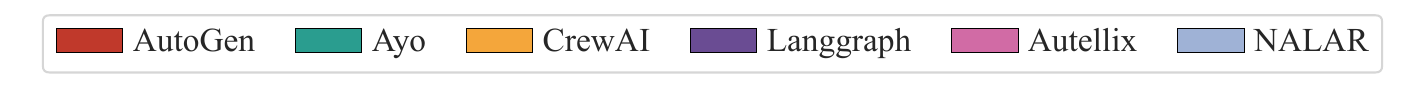}\\
      \vspace{-3pt}
    \begin{subfigure}[b]{0.31\textwidth}
    \includegraphics[width=\textwidth]{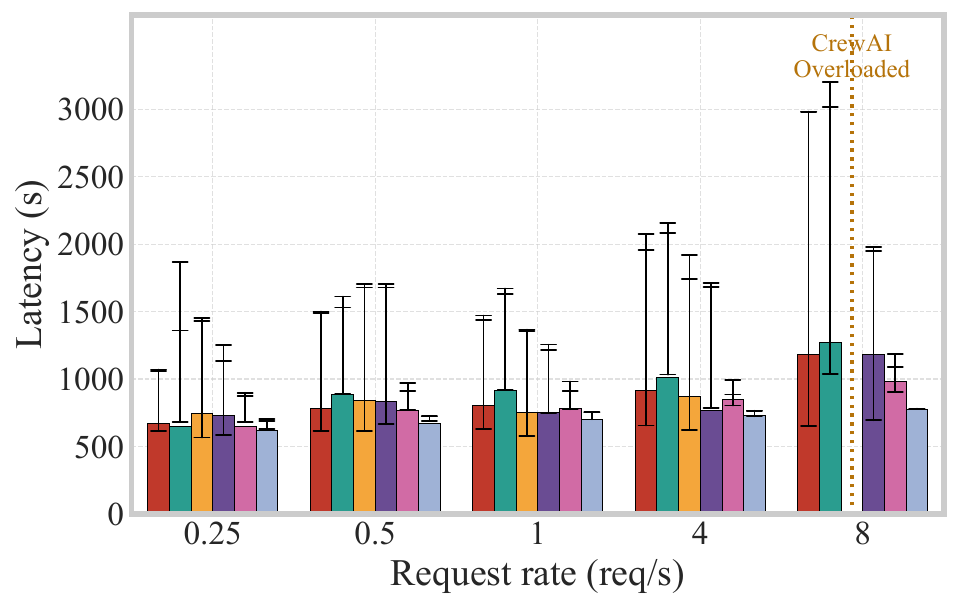}
    \vspace{-12pt}
    \caption{{\textbf{Financial Analyst Agent}}}
    \label{fig:fin_analysis_agent}
    \end{subfigure}
    \centering
    \begin{subfigure}[b]{0.31\textwidth}
    \includegraphics[width=\linewidth]{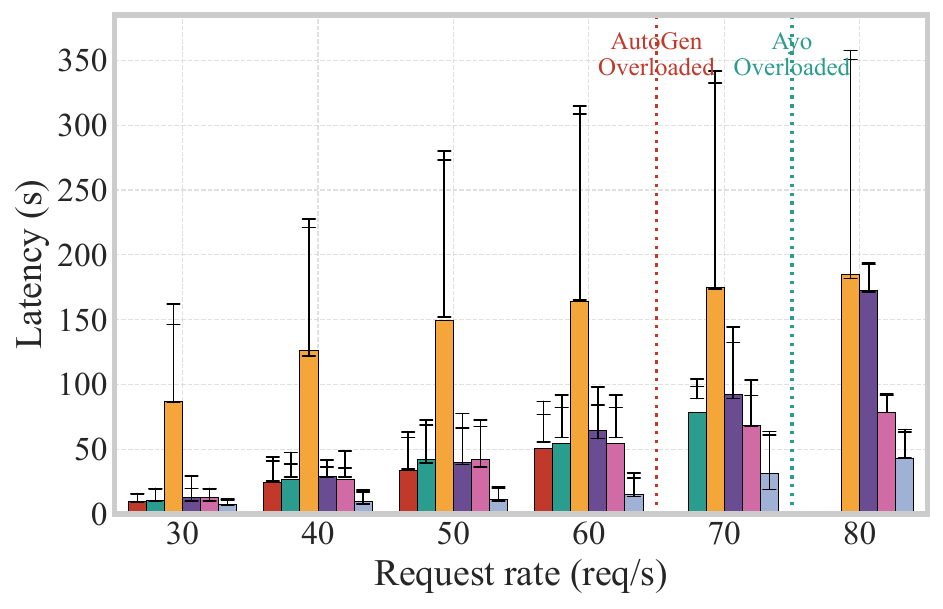}
    \vspace{-12pt}
    \caption{\small{\textbf{Router-based Workflow} }}
    \label{fig:multi-lang} 
    \end{subfigure}
    \centering
    \begin{subfigure}[b]{0.31\textwidth}
    \includegraphics[width=\linewidth]{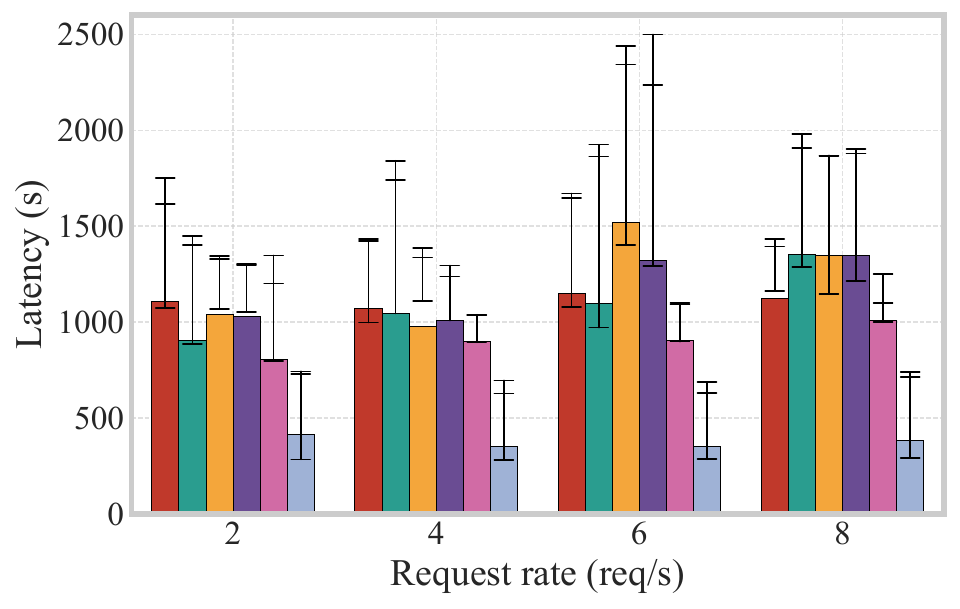}
    \vspace{-12pt}
    \caption{{\textbf{Software Engineering Workflow} }}
    \label{fig:swe-workflow} 
    \end{subfigure}
    \vspace{-10pt}
\caption{\textbf{End-to-End Evaluation} The bars represent average latency, the whiskers represent P50, P95 and P99 latencies.}    
\label{fig:aend-to-end}
\vspace{-10pt}
\end{figure*}

\noindent\textbf{Financial Analyst Workflow.}
Figure~\ref{fig:fin_analysis_agent} evaluates the Financial Analyst workflow. Baselines preserve KV-cache reuse through strict session affinity, which pins each session to its initial GPU and causes head-of-line blocking. By migrating queued futures together with their KV-cache state, \system reduces average latency by up to 35\% and P95--P99 latency by 34--74\%. At 8 RPS, \system reduces P99 latency from over 3,000s to approximately 800s, a $3.75\times$ improvement, while CrewAI fails because of LLM-call timeouts.

% Figure~\ref{fig:fin_analysis_agent} evaluates the Financial Analyst workflow. To maximize KV cache reuse across agents sharing extensive context, baselines enforce strict session affinity, pinning all sequential requests for a given sessionID to their initial GPU. This rigid pinning inevitably induces severe head-of-line blocking. \system eliminates this by treating KV caches as migratable state. 
% The global control plane dynamically migrates queued futures together with their associated KV-cache state across GPUs.
% Consequently, \system improves average latency by up to 35\% over the best baseline. Although massive LLM generation times inherently bound average speedups, \system's impact on the tail is substantial: it reduces P95 and P99 latencies by 34\% to 74\%. At peak load (8 RPS), baselines suffer catastrophic queuing (P99 exceeding 3,000s), whereas \system robustly bounds P99 to approximately 800s (a 3.75$\times$ improvement). The CrewAI baseline failed entirely due to LLM call time-outs, crashing before completing the 8 RPS workload across several attempts. 

\noindent\textbf{Router-based Workflow.}
Figure~\ref{fig:multi-lang} evaluates the router-based workflow, where shifts in query characteristics produce branch-load skew exceeding 90\%~\cite{stojkovic2025dynamollm}. Because existing frameworks use static resource allocations, they strand capacity on underutilized branches while overloaded branches accumulate requests. Consequently, AutoGen and Ayo fail to sustain 70 and 80~RPS, respectively. \system instead monitors branch load and dynamically reallocates compute from underutilized to overloaded component leading to reduction in average latency by up to $3.38\times$, and maintains sub-50s average latency at the peak load of 80~RPS.

\noindent\textbf{Software Engineering Workflow.}
In this highly dynamic workload, \system improves end-to-end latency by up to $2.9\times$. Open-SWE's recursive structure, in which nondeterministic test failures trigger repeated developer invocations, creates both transient load imbalance and head-of-line blocking. \system mitigates these effects through late-binding load balancing and workflow-aware request prioritization, whereas the baselines experience up to $2.1\times$ greater load imbalance. Recursion also causes requests to return repeatedly to the same LLM-backed agent, creating opportunities to reuse inference state. \system's fine-grained KV-cache management preserves this state across retries and reduces the cost of resuming subsequent invocations.

\subsection{Ablation, Fault-tolerance, Overhead and Scalability}
%\vspace{-1em}
\label{sec:ablation}
Next, we perform ablation of \system's contributions.
% understand the overhead introduced and the scalability achieved by \system's design choices. 

% As an academic lab without access to large-scale GPU resources, we follow prior work~\cite{agarwal2024blox, chung2024reducing} and use CPU-based emulation to study \system’s overhead and design implications on scalability. We mimic the SWE workflow, and for LLM inference calls, use the offline profile information to mimic execution behavior. Since \system’s design is not tied to GPUs, we believe the approach is reasonable, and enables us to showcase \system's scalability when deployed in large GPU clusters.
% Where used, we highlight the use of emulation.
%\vspace{-1em}
%\vspace{-1em}
\begin{figure*}[t]
    \centering
    % First minipage
    \begin{minipage}[t]{0.28\linewidth}
    \includegraphics[width=\linewidth]{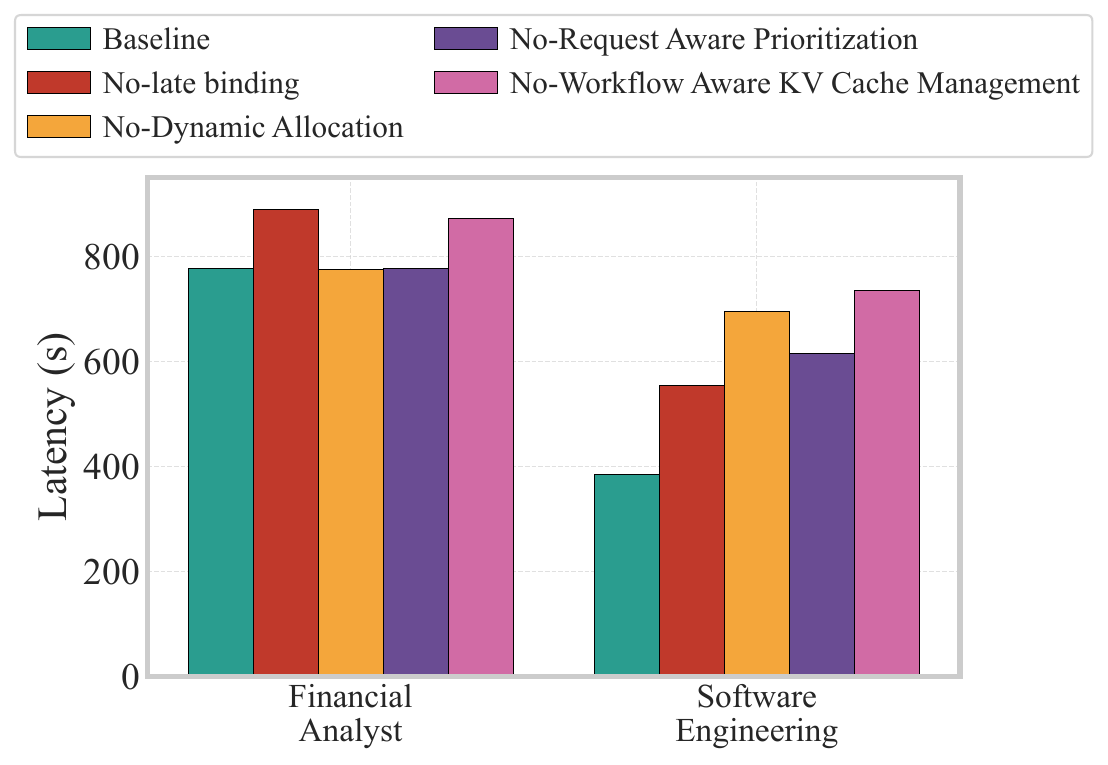}
    \vspace{-10pt}
    \caption{\textbf{Policy Ablation:}  By switching off one mechanism at a time we observe it's impact. To evaluate we pick the highest load reported \S\ref{sec:end-to-end} for these workflows.}
    \label{fig:policy-ablation}
    \end{minipage}
    \hfill
    % Second minipage
    \begin{minipage}[t]{0.28\linewidth}
      \centering
    \includegraphics[width=\linewidth]{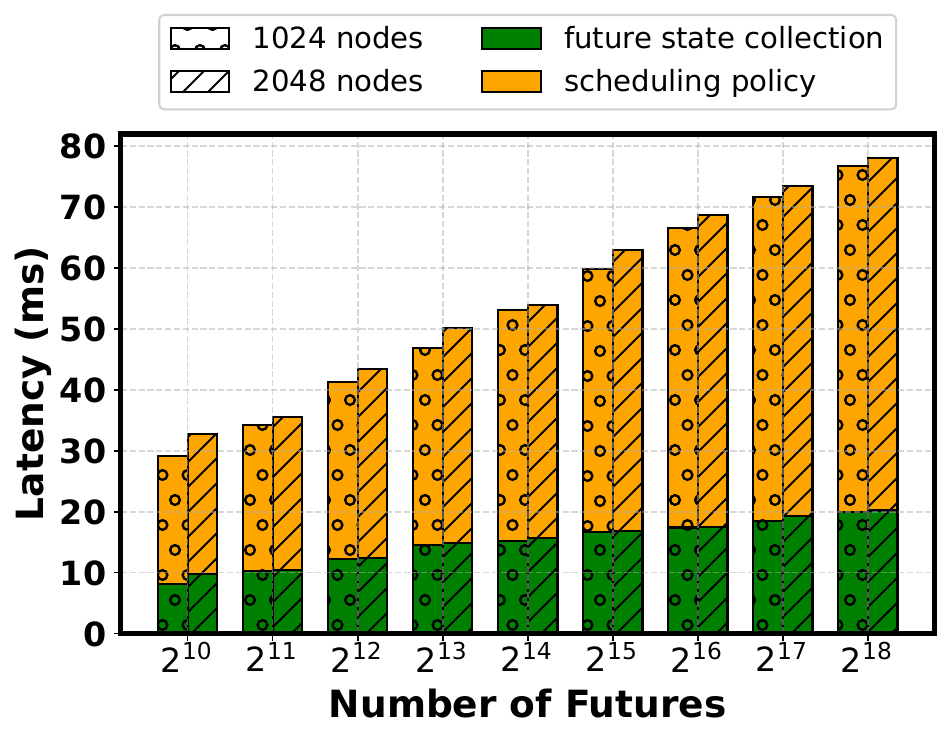}
    \vspace{-10pt}
    \caption{\textbf{Global Control Loop Latency:} Even at a 2048 Node and 262K futures, the loop takes only 78.6ms. Major time (over 80\%) is spent in scheduling policy logic.}
    \label{fig:globa_control_time}
    \end{minipage}\hfill
   \begin{minipage}[t]{0.42\linewidth}
    \includegraphics[width=1\linewidth]{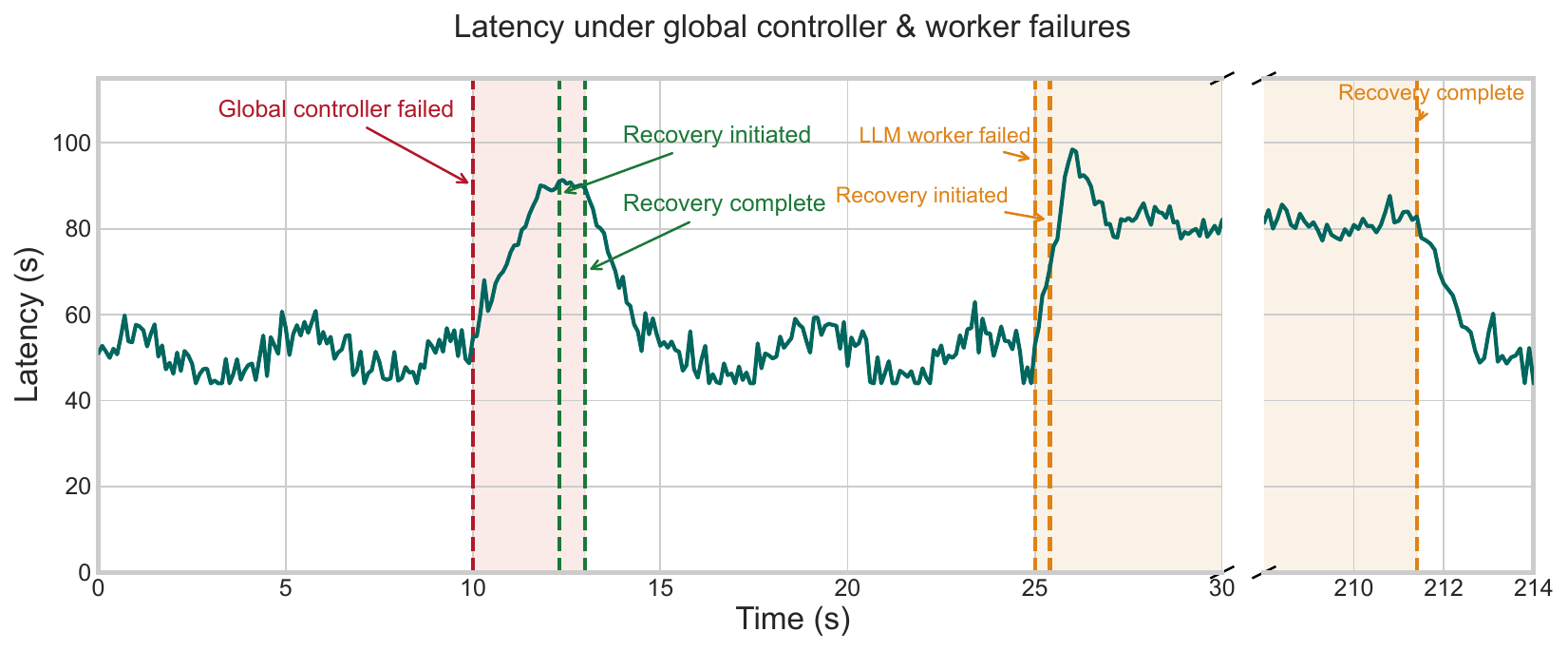}
    \vspace{-10pt}
    \caption{{\textbf{Fault Tolerance:} We study fault tolerance for the router workflows (lowest average latency) despite the global controller failing \system continues serving.}}
    \label{fig:fault-tolerance}
   \end{minipage} 
    \vspace{-10pt}
\end{figure*}

% \begin{figure}
% \vspace{-15pt}
% \end{figure}

\subsubsection{Ablation}

\noindent\textbf{Impact of Policy Mechanisms.}
Figure~\ref{fig:policy-ablation} isolates the contribution of each mechanism in default policy. For Financial Analyst, disabling late-binding load balancing or workflow-aware KV cache management increases latency by 12.3\% and 10.5\%, respectively; the remaining mechanism provide limited benefit for this non-cyclic workflow. In contrast, for the cyclic Software Engineering workflow, disabling late-binding load balancing, dynamic resource allocation, request prioritization, or KV-cache management increases latency by 30.5\%, 80.2\%, 51.3\%, and 83.2\%, respectively. These results show that policy effectiveness depends on workflow structure, with dynamic and cyclic workflows benefiting from coordinated placement, capacity, prioritization, and KV cache management.

\noindent\textbf{Two-level control.}\
\noindent\underline{\emph{Note:}}
Following prior work~\cite{agarwal2024blox, chung2024reducing}, we use CPU-based emulation to evaluate \system at scales beyond our available GPU resources. We emulate the SWE workflow using offline LLM profiles for inference execution and clearly identify all emulated experiments.

We evaluate this in an emulation environment with 1{,}024 CPU nodes and 8{,}192 agents, scaling the number of concurrent futures in \system from $2^{10}$ to $2^{18}$. To isolate control overhead, we enforce an SRTF policy using exact oracle execution times from our profiler.
Table~\ref{tab:ablation-two-level} compares \system's two-level control with a centralized controller. At modest scale ($\le 16$K active futures), centralized routing already adds up to 4\,ms of overhead, dominated by network and queueing delay. This overhead is prohibitive, since roughly $30\%$ of tool invocations complete in under 3\,ms (Figure~\ref{fig:runtime-tool-cdf}). Beyond 16K futures, centralized queueing dominates, increasing scheduling latency to over 72\,ms at 130K futures. By delegating routing to component-level controllers, \system's two-level design avoids centralized bottlenecks.
\subsubsection{Overhead}
%\vspace{-1em}
Distributed RPC, future materialization, and Node Store updates introduce execution overhead. On a sequential, single-node Financial Analyst workload, \system adds 5.4\,ms on average and 6.9\,ms at most over native Python. This overhead is negligible relative to the workload's 300--400\,s end-to-end execution time.

\subsubsection{Scalability}
% \vspace{-1em}
% \noindent\textbf{Scalability with many futures.}
We evaluate the scalability of \system's global control loop in the emulation setup discussed above, scaling active futures up to 262,144 ($2^{18}$) across two cluster configurations: 1,024 nodes (8,192 agents) and 2,048 nodes (16,384 agents). The global controller operates strictly off the critical path; its latency dictates only the propagation speed of policy updates, never gating actual task execution.

As Figure~\ref{fig:globa_control_time} illustrates, when running all four mechanisms in \system's default policy (\S\ref{sec:evaluation}), control loop latency is independent of cluster size and primarily depends on the scalability of the user-implemented scheduling policy. Figure~\ref{fig:globa_control_time} shows that optimized Node Store data layouts keep global state aggregation under 20ms, leaving the $\sim$60ms policy evaluation logic as the primary bottleneck. While automatically accelerating user-defined policies remains future work, the controller's out-of-band architecture ensures \system sustains massive scale without penalizing execution latency.

\subsubsection{Fault Tolerance}
We evaluate the failure-recovery mechanisms described in \S\ref{sec:fault-tolerance}. A global-controller failure increases latency by $1.5\times$ because \system temporarily cannot adapt resource allocations. After the controller's etcd lease expires, a recovery process launches a replacement controller; failure detection and election take 2.3 seconds, followed by 0.7 seconds to reconstruct global state. Throughout recovery, component-level controllers continue serving requests using their last installed policies, demonstrating that \system decouples request execution from global policy computation. We next terminate an LLM worker. The global controller detects the failure within 0.2 seconds and redistributes resources among the remaining workers, reducing latency by 18.3\% relative to the post-failure peak. The vLLM engine restarts in approximately 185 seconds, after which \system returns the recovered worker to the routing set.

\begin{table}[t]
\centering
\caption{Impact of Two-level Control. }
\vspace{-10pt}
\label{tab:ablation-two-level}
\resizebox{0.5\linewidth}{!}{
\begin{tabular}{@{}ccc@{}}
\toprule
\multirow{2}{*}{Number of Futures} & One-Level Design & Two-level Design \\ \cmidrule(l){2-3} 
                                   & Time(ms)         & Time(ms)         \\ \midrule
1024                               & 1.2              & 0.1              \\
2048                               & 2.3              & 0.1              \\
4096                               & 2.8              & 0.2              \\
8192                               & 3.4              & 0.4              \\
16384                              & 3.9              & 0.4              \\
32768                              & 19.4             & 0.3              \\
65536                              & 32.6             & 0.4              \\
131072                             & 72.3             & 0.4              \\ 
262144                             & 179.8            & 0.4              \\ 
\bottomrule
\end{tabular}
}
\vspace{-15pt}
\end{table}

% Our results show that as the number of futures exceeds 16K, concurrent calls to the global controller introduce significant queuing delays, causing severe slowdown.

% Next, we evaluate the choice of two level design on overhead.
% We calculate the overhead of globally central controller would have had if it needed to route the futures manually so as to maintain the SRTF policy.
% We observe that as the number of futures increases beyond 16K futures, which concurrently calls the global controller, there significant slowdown due to queueing delay.

% \vspace{-1em}
\section{Related Work}
% \vspace{-1em}
%We have described several existing agentic inference works in \S\ref{sec:othersystems}. Next, we discuss other related works. 

\noindent
{\bf The Future Abstraction.}
% \setkeys{}
Futures and promises underpin concurrent programming models~ \cite{hewitt1977actors, liskov1988distributed, chatterjee1989futures, bloom2009thorn} and dynamic task runtimes such as CIEL~\cite{murray2011ciel}, Dask~\cite{peters2023parallel}, and Ray~\cite{moritz2018ray}. Ray schedules tasks in response to runtime events, whereas \system additionally performs periodic global reconciliation that revises placement, priority, and state-management decisions using workflow- and cluster-level context. Supporting these capabilities in Ray would require significant extensions to metadata propagation, scheduling, and inference-state management.

\noindent
{\bf Global Control Plane.}
Logically centralized control planes have appeared in several settings~\cite{ghemawat2003google,gog2016firmament, luksa2017kubernetes,moritz2018ray,hindman2011mesos}. \system draws inspiration from this lineage, but differs in its complete decoupling of local component-level controllers from the global controller, an idea borrowed from SDN systems such as B4~\cite{b4:google:sigcomm}. 
% This separation allows \system to override poor local decisions by migrating tasks, making scheduling changes reversible through job migration.

% \aditya{XXX add things not already coevred earlie}
\section{Conclusion}
% \vspace{-1em}
Nalar exposes dynamic workflow structure through semantic futures and combines low-overhead component-level control with periodic global reconciliation. This architecture supports revisable placement, resource allocation, workflow-aware prioritization, and KV-cache management using a common view of workflow progress and resource conditions. Across the evaluated agent workflows, these mechanisms improve end-to-end and tail performance while keeping global control off the critical path

\clearpage
\bibliographystyle{plain}
\bibliography{ref}
\newpage
\appendix
\clearpage
\section{Details of Mechanisms}
\label{sec:details}
\subsection{Late-binding load balancing}
Algorithm~\ref{alg:late-binding} describes \system's late-binding load-balancing policy. For a stateful future, \system preserves session affinity by routing the future to the instance already associated with its session. For a stateless future assigned to source replica $i$, the controller identifies the least-loaded compatible replica $j$. It considers migration only when the ready-queue difference $Q_i-Q_j$ exceeds the imbalance threshold $\alpha$. The controller then estimates the queueing delay at the source and destination, $W_i$ and $W_j$, and the cost $C_{\mathrm{mig}}$ of transferring the future, its unresolved dependencies, and any associated data. The net migration benefit is $\Delta = W_i-W_j-C_{\mathrm{mig}}$. If $\Delta>\beta$, \system redirects unresolved dependencies to $j$ and migrates the future; otherwise, the future remains at $i$. Thus, the policy revises an initial placement only when the observed imbalance is substantial and the predicted queueing-delay reduction exceeds the migration cost by at least $\beta$. We keep $\alpha$ fixed at 20\%, and estimate $\beta$ by accounting for the metadata to move. 

\begin{algorithm}[t]
\caption{Late-Binding Load Balancing}
\label{alg:late-binding}

\begin{algorithmic}[1]

\Require Queued future $f$
\Require Source replica $i$
\Require Available replicas $\mathcal{I}$
\Require Imbalance threshold $\alpha$
\Require Minimum migration benefit $\beta$

\If{\Call{IsStateful}{$f$}}

    \State $j \gets
    \Call{SessionReplica}{f.\textit{sessionID}}$

    \State \Return $j$

\EndIf

\State $j \gets
\arg\min_{k \in \mathcal{I}} Q_k$

\If{$Q_i - Q_j \leq \alpha$}

    \State \Return $i$

\EndIf

\State $W_i \gets
\Call{QueueDelay}{f,i}$

\State $W_j \gets
\Call{QueueDelay}{f,j}$

\State $C_{\mathrm{mig}} \gets
\Call{MigrationCost}{f,i,j}$

\State $\Delta \gets
W_i - W_j - C_{\mathrm{mig}}$

\If{$\Delta > \beta$}

    \State \Call{RedirectDependencies}{f,j}

    \State \Call{Migrate}{f,i,j}

    \State \Return $j$

\Else

    \State \Return $i$

\EndIf

\end{algorithmic}

\end{algorithm}

\subsection{Dynamic Resource Allocation}
\label{sec:dynamic-resource}

\begin{algorithm}[t]
\caption{Workflow-Aware Resource Reallocation}
\label{alg:block}
\begin{algorithmic}[1]

\Require Runtime statistics $\mathcal{M}_t$
\Require Current allocation $R$
\Require Allocation constraints $C$
\Require Persistence threshold $K = 4$
\Require Minimum throughput gain $\epsilon$
\Require Numerical floor $\delta$

\State $G_t \gets
\Call{FlattenPaths}
{\mathcal{M}_t.\textit{requestPaths}}$

\ForAll{components $v \in G_t$}

    \State $p_v \gets
    \Call{MeanVisitsPerRequest}
    {v, G_t}$

    \State $\bar{\mu}_v \gets
    \Call{PerReplicaCapacity}
    {\mathcal{M}_t[v]}$

    \State $\mu_v \gets
    R_v \cdot \bar{\mu}_v$

    \State $x_v \gets
    \mu_v / \max(p_v, \delta)$

\EndFor

\State $v^* \gets
\arg\min_{v \in G_t} x_v$
\Comment{Estimated bottleneck}

\If{$Q_{v^*}^{\mathrm{ready}} = 0$}

    \State \Return \textsc{NoReallocation}

\EndIf

\If{\Call{Persistent}{$v^*, K$} = \textbf{false}}

    \State \Return \textsc{NoReallocation}

\EndIf

\State $X \gets
\min_{v \in G_t} x_v$

\State $u^* \gets \bot$

\State $X^* \gets X$

\ForAll{components $u \in G_t$ such that $u \neq v^*$}

    \If{$R_u > C.\textit{minInstances}[u]$
    \textbf{and}
    $Q_u^{\mathrm{ready}} = 0$}

        \State $R' \gets R$

        \State $R'_u \gets R_u - 1$

        \State $R'_{v^*} \gets R_{v^*} + 1$

        \If{\Call{Feasible}{$R', C$}}

            \State $X' \gets
            \Call{PredictedThroughput}
            {G_t, \bar{\mu}, R', \delta}$

            \If{$X' > X^*$}

                \State $u^* \gets u$

                \State $X^* \gets X'$

            \EndIf

        \EndIf

    \EndIf

\EndFor

\If{$u^* \neq \bot$
\textbf{and}
$X^* > (1+\epsilon)X$}

    \State \Return
    \Call{Reallocate}
    {$u^*, v^*$}

\Else

    \State \Return
    \textsc{NoReallocation}

\EndIf

\end{algorithmic}

\end{algorithm}
Algorithm~\ref{alg:block} describes \system's workflow-aware resource reallocation policy. At each control interval, \system constructs a flattened graph from the request paths observed in the previous interval. For each component $v$, it computes the mean visits per workflow request, $p_v$, and estimates the component's total service capacity as $\mu_v = R_v\bar{\mu}_v$, where $R_v$ is its current replica count and $\bar{\mu}_v$ is its measured per-replica capacity. The normalized capacity $x_v=\mu_v/\max(p_v,\delta)$ estimates the workflow throughput that component $v$ can sustain. The component with the smallest $x_v$ is the candidate bottleneck. \system reallocates resources only if this component has ready work and remains the bottleneck for $K$ consecutive control intervals, thereby avoiding reactions to dependency stalls or transient load fluctuations. It then considers transferring one replica from each idle component above its minimum allocation to the bottleneck. Candidates that violate the instance-count or cluster-resource constraints in $C$ are discarded. Among the feasible candidates, \system selects the allocation with the highest predicted workflow throughput and applies it only if the improvement exceeds the threshold $\epsilon$.

For each component $v$, \system estimates its mean per-replica service
capacity as

\[
\bar{\mu}_v =
\frac{
    N_v^{\mathrm{busy}}
}{
    A_v^{\mathrm{busy}}
},
\]

where $N_v^{\mathrm{busy}}$ is the number of invocations completed while
runnable work was available, and $A_v^{\mathrm{busy}}$ is the aggregate
busy time of its replicas, measured in replica-seconds. Under allocation
$R$, the component's total effective capacity is

\[
\mu_v = R_v \bar{\mu}_v.
\]

From the flattened execution graph, \system computes $p_v$, the mean
number of visits to component $v$ per workflow request. The corresponding
workflow-normalized capacity is

\[
x_v =
\frac{
    \mu_v
}{
    \max(p_v,\delta)
},
\]

where $\delta$ is a small numerical floor. Thus, $x_v$ estimates the
maximum workflow throughput that component $v$ can sustain under its
current allocation, and the component with the smallest $x_v$ is the
candidate bottleneck.

For a candidate allocation $R'$, \system predicts workflow throughput as

\[
\widehat{X}(R') =
\min_{v \in G_t}
\frac{
    R'_v \bar{\mu}_v
}{
    \max(p_v,\delta)
}.
\]

\system considers only candidate allocations that satisfy the
instance-count and cluster-resource constraints in $C$. Among these
candidates, it selects the allocation with the largest predicted
throughput and performs the reallocation only if

\[
\widehat{X}(R') >
(1+\epsilon)\widehat{X}(R),
\]

where $\epsilon$ is the minimum required throughput improvement.

\begin{algorithm}[t]

\caption{Workflow-Aware KV-Cache Management}

\label{alg:kv-cache-management}

\begin{algorithmic}[1]

\Require Active sessions $\mathcal{S}$

\Require Available GPU memory $M_{\mathrm{free}}$

\Require Required GPU-memory reserve $M_{\mathrm{target}}$

\Require Transfer safety margin $\delta$

\ForAll{sessions $s \in \mathcal{S}$}

    \If{\Call{SessionComplete}{$s$}}

        \State \Call{InvalidateCache}{$s$}

        \State \textbf{continue}

    \EndIf

    \If{\Call{CacheOffloaded}{$s$}}

        \State $T_{\mathrm{ready}} \gets
        \Call{TimeUntilReady}{s}$

        \State $T_{\mathrm{restore}} \gets
        \Call{RestoreTime}{s}$

        \If{$T_{\mathrm{ready}}
        \leq
        T_{\mathrm{restore}} + \delta$}

            \State \Call{PrefetchCacheAsync}{$s$}

        \EndIf

    \EndIf

\EndFor

\If{$M_{\mathrm{free}} \geq M_{\mathrm{target}}$}

    \State \Return

\EndIf

\State $\mathcal{C} \gets
\Call{SuspendedSessions}{\mathcal{S}}$

\ForAll{sessions $s \in \mathcal{C}$}

    \State $T_{\mathrm{idle}}[s] \gets
    \Call{TimeUntilReady}{s}$

    \State $T_{\mathrm{move}}[s] \gets
    \Call{OffloadTime}{s}$
    \Statex \hspace{\algorithmicindent}
    $+
    \Call{RestoreTime}{s}$

\EndFor

\State $\mathcal{C} \gets
\Call{SortDescending}
{\mathcal{C}, T_{\mathrm{idle}}}$

\ForAll{sessions $s \in \mathcal{C}$}

    \If{$T_{\mathrm{idle}}[s]
    >
    T_{\mathrm{move}}[s] + \delta$}

        \State \Call{OffloadCacheIncrementally}{$s$}

        \State $M_{\mathrm{free}} \gets
        M_{\mathrm{free}}
        +
        \Call{CacheSize}{s}$

    \EndIf

    \If{$M_{\mathrm{free}} \geq M_{\mathrm{target}}$}

        \State \textbf{break}

    \EndIf

\EndFor

\end{algorithmic}

\end{algorithm}

\ subsection {Workflow-aware KV cache managment}a
\label{ref:kv-cache-managment}

Algorithm~\ref{alg:kv-cache-management} describes \system's
workflow-aware KV-cache policy. The global controller first invalidates
the caches of completed sessions. For an offloaded session, it initiates
asynchronous prefetching when the predicted time until the session becomes
runnable falls below the cache-restoration time, including a safety margin
$\delta$. Under GPU memory pressure, \system ranks suspended sessions by
their predicted remaining idle time and incrementally offloads caches
whose idle intervals exceed the combined offload and restoration costs.
Offloading continues only until the desired GPU-memory reserve is restored.

For each suspended session $s$, \system estimates the time until the
session becomes runnable as

\[
\widehat{T}_{\mathrm{ready}}(s)
=
\max_{f \in \mathcal{D}(s)}
\widehat{T}_{\mathrm{remaining}}(f),
\]

where $\mathcal{D}(s)$ contains the unresolved dependencies that must
all complete before session $s$ can resume. For each dependency $f$,
\system estimates its remaining time using the historical mean execution
time of futures at the corresponding component:

\[
\widehat{T}_{\mathrm{remaining}}(f)
=
\max
\left(
0,
\overline{T}_{\mathrm{component}(f)}
-
T_{\mathrm{elapsed}}(f)
\right).
\]

\section{Telemetry Data Collected.}
Table~\ref{tab:telemetry} provides detailed information about telemetry provided by local controllers and aggregated by the global controller.

% \begin{table}[t] \centering \caption{ Telemetry collected by \system's component-level controllers. } \label{tab:telemetry} \begin{tabular} { @{} p{0.28\columnwidth} p{0.64\columnwidth} @{} } \toprule \textbf{Telemetry} & \textbf{Description} \\ \midrule Identifiers & Session and future IDs associated with each invocation. \\ Workflow metadata & Invoked component, creator, executor, dependencies, and consumers. \\ Lifecycle timestamps & Future creation, readiness, execution start, and completion times. \\ Latency & Dependency-wait, ready-queue, and execution times. \\ Instance load & Numbers of blocked, ready, running, and outstanding futures. \\ Resource state & Allocated resources, available capacity, and utilization. \\ \bottomrule \end{tabular} \end{table}

\begin{table*}[t] \centering \caption{\system telemetry collected by component-level controllers. Per-future telemetry captures workflow dependencies and timing, while instance-level metrics summarize local load and resource availability.} \label{tab:telemetry} 
\small \begin{tabular}{p{0.22\columnwidth} p{0.20\columnwidth} p{0.50\columnwidth}} \toprule \textbf{Category} & \textbf{Metric} & \textbf{Description} \\ \midrule \multirow{6}{*}{Future metadata} & Session ID & UniqueID associated with a session. \\ & Future ID & Unique ID associated \\ & Component & Identifies the agent or tool invoked by the future. \\ & Creator & Component that created the future. \\ & Executor & Agent or tool instance assigned to execute the future. \\ & Dependencies & Futures whose values must be available before the future becomes ready. \\ & Consumers & Components waiting to consume the future's result. \\ \midrule \multirow{5}{*}{Future timing} & Creation time & Time at which the invocation is intercepted and the future is created. \\ & Ready time & Time at which all dependencies have resolved and the future becomes runnable. \\ & Start time & Time at which execution begins on the assigned component instance. \\ & Completion time & Time at which execution completes and the result becomes available. \\ & Dependency-wait time & Time from future creation until all dependencies resolve. \\ & Queueing time & Time spent in the ready queue before execution begins. \\ & Execution time & Time spent executing on the assigned component instance. \\ \midrule \multirow{4}{*}{Instance load} & Ready futures & Number of dependency-resolved futures waiting to execute. \\ & Blocked futures & Number of futures waiting for one or more dependencies to resolve. \\ & Running futures & Number of futures currently executing on the instance. \\ & Outstanding futures & Total number of incomplete futures assigned to the instance. \\ \midrule \multirow{3}{*}{Resource state} & Allocated resources & Compute and memory resources allocated to the component instance. \\ & Available resources & Compute and memory capacity currently available for additional work. \\ & Utilization & Observed utilization of the instance's allocated compute and memory resources. \\ \bottomrule \end{tabular} \end{table*}

\section{Controller interaction}
\label{sec:controller-interaction}
We illustrate how the global and component-level controllers coordinate to migrate a future. Consider a workflow in which the driver invokes \texttt{agentA}, whose output is subsequently consumed by \texttt{agentB}. Suppose two instances of \texttt{agentB}, \texttt{B:0} and \texttt{B:1}, are available. Invoking \texttt{agentA} creates future \texttt{f1}, and subsequently invoking \texttt{agentB} creates future \texttt{f2}, with \texttt{f1} as its dependency. Both futures are created by the driver. Initially, \texttt{B:0} is assigned as the executor of \texttt{f2}. Consequently, \texttt{B:0} is registered as a consumer of \texttt{f1}, while the driver is registered as a consumer of \texttt{f2}. Suppose the global controller decides to migrate the queued future \texttt{f2} from \texttt{B:0} to \texttt{B:1}. The migration proceeds in six steps. In Step~\Circled{1}, the global controller sends a \texttt{migrate} command to the component-level controller at \texttt{B:0}. Upon receiving the command, \texttt{B:0} marks \texttt{f2} as migrating, preventing it from being activated locally. In Step~\Circled{2}, \texttt{B:0} contacts \texttt{A:0}, the component-level controller that produced \texttt{f1}, to determine whether its value has already been delivered or is in transit. If \texttt{f1} has not yet been transmitted, \texttt{A:0} replaces \texttt{B:0} with \texttt{B:1} in \texttt{f1}'s consumer set (Step~\Circled{3}), causing the dependency value to be delivered directly to the destination. If the value has already been delivered to \texttt{B:0}, \texttt{B:0} forwards it to \texttt{B:1}. If the value is in transit, \texttt{B:0} waits for its arrival and then forwards it. These cases ensure that \texttt{B:1} receives the dependency required to execute \texttt{f2}. Once dependency delivery has been resolved, the component-level controllers update \texttt{f2}'s executor from \texttt{B:0} to \texttt{B:1} in Step~\Circled{4}; its creator remains the driver. In Step~\Circled{5}, \system transfers any KV-cache state associated with f2, when applicable. After completing the dependency and state transfers, \texttt{B:0} removes \texttt{f2} from its local queue and acknowledges the handoff to \texttt{B:1}. Only after receiving this acknowledgment does \texttt{B:1} activate \texttt{f2} (Step~\Circled{6}).
 \begin{figure*}[t]
    \begin{center}
    \begin{subfigure}[b]{0.33\textwidth}
        \includegraphics[width=\textwidth]{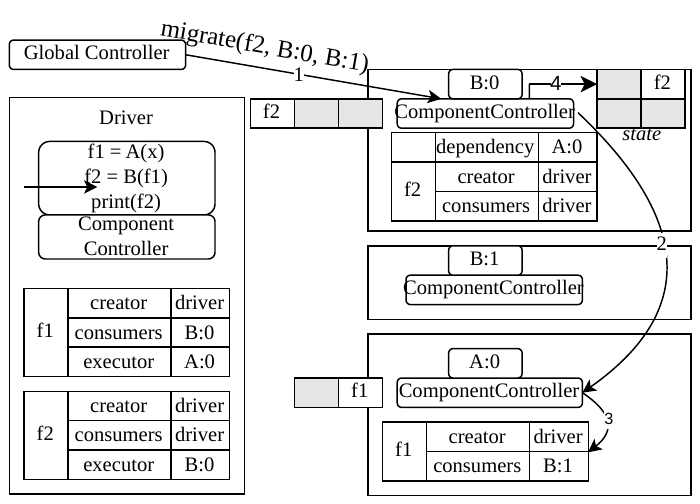}
    \vspace{-10pt}
    \caption{\small{\textbf{Migration initiation}}}
    \label{fig:future_creation}
    \end{subfigure}
    \begin{subfigure}[b]{0.33\textwidth}
    \includegraphics[width=\textwidth]{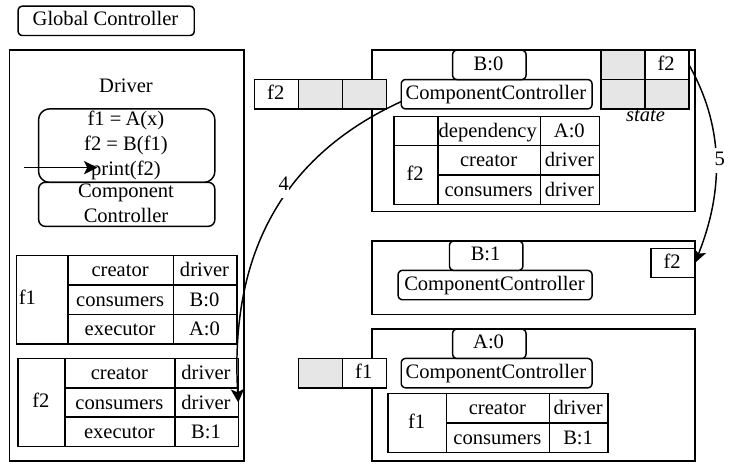}
    \vspace{-10pt}
    \caption{\small{\textbf{Dependency Updates}}}
    \label{fig:future_update}
    \end{subfigure}
    % \vspace{-0.2in}
    % \label{fig:future_management}
    \begin{subfigure}[b]{0.33\textwidth}
    \includegraphics[width=\textwidth]{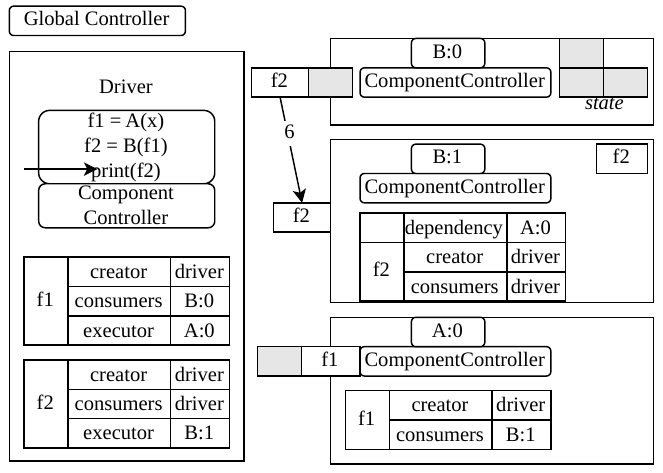}
    \vspace{-10pt}
    \caption{\small{\textbf{Migration Completion}}}
    \label{fig:future_complete}
    \end{subfigure}
    \vspace{-15pt}
    \caption{{\small\textbf{Control Interaction in \system.} The above figure shows the interaction and relevant updates to metadata when a future is being migrated. An important feature is that it's entirely locally coordinated, \ie the global controller only issues the migrate command, the component level controllers coordinate it among themselves. }}
    \label{fig:future_management}
    \end{center}
    % \vspace{-20pt}
\end{figure*}

\section{Adding New Policies}
\label{sec:more_policies}
% \vspace{-1em}
Next, we show how \system's scheduling API allows developers to easily implement diverse and effective policies.

\noindent
{\bf Minimize JCT.}
To minimize avg job completion time (JCT), schedulers traditionally employ Shortest Remaining Time First (SRTF) policies. Unlike the default progress-based policy in 6.2,
we approximate SRTF using a topological heuristic that prioritizes
invocations originating from the deepest stages of the execution graph.
% A common way to reduce job completion time is to prioritize jobs with the least remaining work, \ie Shortest Remaining Time First (SRTF). In call-graph–structured workloads such as the financial analyst agent, a practical heuristic to improve average completion latencies is to prioritize calls originating from later stages of the graph. 
Implementing this policy in \system requires just \emph{14 lines} of Python running on the global controller. 
The policy can be concisely expressed due to the well-designed policy interface provided by \system.
 % \aditya{briefly remind the reader what in your systems enables this}
We observe that the policy reduces average JCT by over 2.4\% at the cost of a 3.3\% increase in P95 latency. 

\noindent\textbf{Control Makespan.}
To reduce the makespan compared to default approaches like FCFS we can prioritize the Longest Processing Time (LPT) job first. In call-graph workflows such as software engineering, this corresponds to prioritizing jobs that re-enter the graph because they failed to meet the specification. Implementing this policy also required just \emph{12 lines} of code. We observed that it reduced makespan by 5.8\%, with a 2.6\% increase in P95 latency for Software Eng workflow.

% \noindent\underline{\emph{Takeaway:}} 
Although performance gains are modest, \system demonstrably simplifies the exploration of custom scheduling policies for agentic inference. Implementing comparable logic in AutoGen—our strongest baseline—proved infeasible; its asynchronous messaging architecture lacks the requisite fine-grained control primitives.

\section{Impact of Hints}
\label{sec:impact-hints}
% \begin{figure}
%     \centering
%     \includegraphics[width=0.7\linewidth]{figures/hint-ablation.pdf}
%     \caption{\small\textbf{Impact of Hints:} For a popular Text-2-sql workflow\cite{li2026dpc}, where sandboxes can be utilized across multiple candidates, where queries can reuse sandboxes, \system improves average latency by over 10.6$\times$.}
%     \label{fig:hint-use}
% \end{figure}
% We demonstrate the utility of \system's hint interface using the text-to-SQL workflow from DPC~\cite{li2026dpc}. The workflow translates a natural-language request into SQL by generating several candidate queries and evaluating them on a sampled database containing tens of gigabytes of data. By default, each request within a session launches a new sandbox, copies a database checkpoint, and initializes an evaluation harness. Because the dataset and sandbox configuration remain unchanged across requests within a session, the developer can mark the evaluation component as \emph{reusable}, allowing \system to retain and reuse the initialized environment. Figure~\ref{fig:hint-use} shows that this single-line hint reduces provisioning and data-copying time by 63\%. Without the hint, repeated environment initialization becomes a bottleneck as the request rate increases, causing rapidly growing queueing delay. Reusing the initialized sandbox eliminates this redundant work and stabilizes performance.

We evaluate \system's hint interface using the text-to-SQL workflow from DPC~\cite{li2026dpc}. For each natural-language request, the workflow generates several candidate SQL queries and evaluates them on a sampled database containing tens of gigabytes of data. By default, each candidate evaluation within a session launches a new
sandbox, copies a database checkpoint, and initializes an evaluation
harness. Because the dataset and sandbox configuration remain unchanged
across candidate evaluations within the session, the developer can mark
the evaluation component as reusable, allowing \system to retain and
reuse the initialized environment.
 Across offered loads of 0.01, 0.02, 0.04, 0.08, and 0.12 requests/s, this single-line hint reduces provisioning and data-copying time by 63\%. At higher loads, eliminating this repeated setup reduces average end-to-end latency by up to $10.6\times$. The end-to-end improvement is larger than the reduction in setup time because, without reuse, sandbox provisioning becomes a serialized bottleneck amplifying queueing delay as load increases. Reusing the initialized environment removes this bottleneck and maintains substantially lower latency across the evaluated request rates. The \emph{reusable} hint is a developer-provided semantic guarantee: the developer is responsible for ensuring that the component can be safely reused within the declared scope. \system does not infer reusability or verify application-level isolation; it uses the hint only to manage the component's placement and lifetime.

 \section{Stub-Generator Limitation}
    \label{sec:stub-limitation}
    Stub Generator Limitations. The stub generator requires
component interfaces to be statically identifiable. It there-
fore cannot export methods created through dynamic meta-
programing such as runtime reflection and monkey-patching,
nor does it support native OS-level concurrency primitives
like process forking. It does, however, seamlessly interoper-
ate with standard asynchronous coroutines (async/await).
These restrictions provide stable component boundaries over
which Nalar can reliably track and control distributed in-
vocations
\end{document}